\documentclass[fleqn,usenatbib]{mnras}

\usepackage{newtxtext,newtxmath}
\usepackage[T1]{fontenc}

\DeclareRobustCommand{\VAN}[3]{#2}
\let\VANthebibliography\thebibliography
\def\thebibliography{\DeclareRobustCommand{\VAN}[3]{##3}\VANthebibliography}

\usepackage{graphicx}    % Including figure files
\usepackage{amsmath}    % Advanced maths commands
\usepackage{tabulary}
\usepackage{footnote}
\usepackage{hyperref}
\usepackage{natbib}
\usepackage[dvipsnames]{xcolor}
\bibpunct{(}{)}{;}{a}{}{,}

\newcommand{\approximately}{\mathord{\sim}}

\title[Galaxy Manifold]{Galaxy Manifold: Characterizing and understanding galaxies with two parameters}

\author[S. Cooray et al.]{Suchetha Cooray,$^{1}$\thanks{E-mail: \href{mailto:cooray@nagoya-u.jp}{cooray@nagoya-u.jp}} $^{,\thanks{JSPS Fellow (DC1)}}$
Tsutomu T. Takeuchi,$^{1, 2}$
Daichi Kashino,$^{3, 1}$\newauthor
Shuntaro A. Yoshida,$^{1}$
Hai-Xia Ma,$^{1}$ and
Kai T. Kono$^{1}$
\\
\\
$^{1}$ Division of Particle and Astrophysical Science, Nagoya University, Furo-cho, Chikusa-ku, Nagoya 464–8602, Japan\\
$^{2}$ The Research Center for Statistical Machine Learning, The Institute of Statistical Mathematics, 
10-3 Midori-cho, Tachikawa, Tokyo 190-8562, Japan\\
$^{3}$ Institute for Advanced Research, Nagoya University, Furo-cho, Chikusa-ku, Nagoya 464–8602, Japan
}

\date{Accepted XXX. Received YYY; in original form ZZZ}

\pubyear{2022}

\begin{document}
\label{firstpage}
\pagerange{\pageref{firstpage}--\pageref{lastpage}}
\maketitle

\begin{abstract}

    We report the discovery of a two-dimensional Galaxy Manifold within the multi-dimensional luminosity space of local galaxies. The multi-dimensional luminosity space is constructed using 11 bands that span from far ultraviolet to near-infrared for redshift < 0.1 galaxies observed with \textsl{GALEX}, SDSS, and UKIDSS. The two latent parameters are sufficient to express 93.2\% of the variance in the galaxy sample, suggesting that this Galaxy Manifold is one of the most efficient representations of galaxies. The transformation between the observed luminosities and the manifold parameters as an analytic mapping is provided. The manifold representation provides accurate (85\%) morphological classifications with a simple linear boundary, and galaxy properties can be estimated with minimal scatter (0.12 dex and 0.04 dex for star formation rate and stellar mass, respectively) by calibrating with the two-dimensional manifold location. Under the assumption that the manifold expresses the possible parameter space of galaxies, the evolution on the manifold is considered. We find that constant and exponentially decreasing star formation histories form almost orthogonal modes of evolution on the manifold. Through these simple models, we understand that the two modes are closely related to gas content, which suggests the close relationship of the manifold to gas accretion. Without assuming a star formation history, a gas-regulated model reproduces an exponentially declining star formation history with a timescale of $\sim$1.2 Gyrs on the manifold. Lastly, the found manifold suggests a paradigm where galaxies are characterized by their mass/scale and specific SFR, which agrees with previous studies of dimensionality reduction. 
   
\end{abstract}

\begin{keywords}
methods: data analysis -- galaxies: evolution -- galaxies: fundamental parameters –- galaxies: statistics -- galaxies: star formation -- galaxies: stellar content 
\end{keywords}

\section{Introduction} \label{sec:introduction}

    A key issue in observational astronomy is understanding how galaxies evolve in their lifetimes. However, galaxy evolution is a complicated interplay of many multiscale processes. Some physical processes that we know of are accretion of gas into the haloes \citep[e.g.,][]{Rubin_2012,Somerville_2015,Zabl_2019}, cooling of gas to form stars \citep[e.g.,][]{Ribau_2011,Saintonage_2011a,Tacconi_2013}, feedback in that interfere with star formation \citep[e.g.,][]{Springel_2005,Fabian_2012,Tombesi_2015,Takeuchi_2022}, and galaxy merging \citep[e.g.,][]{Moster_2011,Hopkins_2013,Fensch_2017}. 
    
    To understand these processes, astronomers can observe galaxies' spectral energy distribution (SEDs), which encodes information about these various processes. Complex physical models can now produce the SEDs that closely mimic observed SEDs \citep[e.g.,][]{Maraston_2011,Conroy_2013,Nishida_2022} and these synthetic SEDs can be used to fit the observations to derive physical properties of galaxies such as star formation rates (SFR), stellar mass ($M_*$), dust attenuation, and star formation history \citep[e.g.,][]{Tojeiro_2007,Leja_2017,Carnall_2018, Robotham_2020}. 
    
    Physical properties derived from observations can give us clues on various aspects of galaxy evolution. Interestingly, many of these properties show clear correlations between them, and many empirical relations called scaling relations have been discussed \citep[e.g.,][]{Faber_1973, Kormendy_1977, Dressler_1987, Roberts_1994, Kennicutt_1998, Wijesinghe_2012}. Many scaling relations have been joined in 3D to form 2D fundamental planes \citep[e.g.,][]{Djorgovski_1987, Jeong_2009, Lara-Lopez_2010, Bezanson_2013, Porter_2014, Kashino_2019}. Such fundamental planes suggest that galaxies lie in an intrinsically low dimensional space within the higher dimensional space, which can also be called a manifold. The manifold where galaxies distribute themselves within the higher dimensional space can be called a "Galaxy Manifold" and has been discussed in the past \citep{Djorgovski_1992}. A continuous galaxy manifold representing the fundamental parameters will greatly facilitate understanding the general properties of galaxy evolution.
    
    Such a galaxy manifold could be found in a data-driven way by using dimensionality reduction, which can also be considered a form of manifold learning in this context. One of the most popular dimensionality reduction techniques is principal component analysis (PCA). For example, \citet{Yip_2004a} used PCA to find components that best approximate linear combinations to the SDSS spectra and found that 50 components are necessary to acceptably represent the data. To combat the issue of having too many components in the latent space with linear methods, non-linear dimensionality reduction techniques have also been used on observations \citep[e.g.,][]{Vanderplas_2009,in_der_Au_2012,Rahmani_2018,Hemmati_2019, Portillo_2020,Davidzon_2022}.
    
    However, a significant drawback of these powerful non-linear techniques is that the transformations between the latent and data space are complicated and cannot be written down analytically. A different solution to the above can be to choose the input features for dimensionality reduction more wisely. Astronomers have traditionally excelled in this process through photometry. Photometric filters are designed to be the most informative of galaxies, as spectroscopy is too expensive. Therefore, looking for the manifold within the multi-dimensional luminosity space is meaningful. 
    
    Galaxies distribute bimodally in the color and magnitude space, where we have the star-forming "blue cloud" and the more quiescent "red sequence" \citep[e.g.,][]{Tully_1982,Strateva_2001,Baldry_2004,Baldry_2006}. The transitional population is considered the "green valley" \citep[e.g.,][]{Bell_2004, Faber_2007, Martin_2007, Schiminovich_2007, Wyder_2007, Mendez_2011, Goncalves_2012}. There have also been works questioning this view, where galaxies are defined on a curve within the multi-dimensional space \citep{Ascasibar_2011}. A manifold representing the galaxy's evolution as a continuous sequence would be advantageous as it would better represent the evolutionary stage and its physical parameters. As a solution, we consider near-ultraviolet (NUV) bands in our analysis as galaxies are known to distribute continuously from "blue cloud" to "red sequence" when represented with NUV-based colors \citep[e.g.,][]{Bundy_2010, Chilingarian_2012, Arnouts_2013, Cibinel_2013, Davidzon_2016, Siudek_2018}.
    
    In this work, we report the discovery of a 2D galaxy manifold within the multi-dimensional luminosity space from far ultraviolet to near-infrared, which can be easily recovered with analytical transformations. The found manifold can be considered the ideal representation of the galaxy distribution in the color space and provides a convenient tool to characterize galaxies. In Sections \ref{sec:data} and \ref{sec:analysis}, we explain the data and the methods used to find this manifold. After that, we discuss the relationship between the found manifold and the galaxy's physical processes in Section \ref{sec:physical_properties}. In Section \ref{sec:evolution}, we consider the scenario where galaxies evolve on the manifold using simple models of galaxy evolution. Section \ref{sec:discussion} includes some discussion on the physical meaning of the manifold axes, drawbacks of the current methods, and some possibilities when using the manifold. We make some conclusions in Section \ref{sec:conclusion}. The paper uses magnitudes expressed in the AB system \citep{Oke_1983} and assumes an universal Chabrier IMF \citep{Chabrier_2003} and Plank 2018 cosmology \citep{Plank_2020}.

\section{Data} \label{sec:data}

    We make use of the Reference Catalog of galaxy Spectral Energy Distributions \citep[RCSED;][]{Chilingarian_2017} for this study. The RCSED catalog is a value-added catalog of $\approximately$800,000 SDSS galaxies, which contains spectra and $K$-corrected photometry. This work employs the integrated photometry available in 11 bands at ultraviolet (UV), optical, and near-infrared (IR) from \textsl{GALEX} (\textit{FUV} and \textit{NUV}), SDSS (\textit{u, g, r, i, z}), and UKIDSS (\textit{Y, J, H, K}), respectively. 

    Out of the whole sample, we limit the sample to galaxies with measurements at all 11 bands, which gives us 90,565 galaxies. We remove galaxies with redshift confidence $\le0.5$ ($\approximately$100 galaxies), which brings down the sample to 90,460. The main reason for the significant difference in number from the parent sample is the small intersecting footprint of the UKIDSS sample. Since we are interested in a universal relation of galaxies, the above sample was volume limited at SDSS \textit{g}-band, giving us a final sample of 27,056 galaxies. This sample was obtained by using a flux limiting curve with ${\rm m}_{\rm AB,\, g} = 18.3$ and maximizing the number of galaxies in the final sample. The optimized redshift and absolute magnitude limits are ($z_{\rm limit}$, $M_{\rm limit}$) = (0.097, -20.016). 

\section{Finding the Galaxy Manifold} \label{sec:analysis}

    The above galaxy sample with 11 absolute magnitudes (features) is considered for dimensionality reduction. Dimensionality reduction transforms a high-dimensional data matrix into a lower-dimensional space while retaining the most meaningful characteristics of the original data matrix. We will reduce the number of dimensions from 11 to a much smaller number of components while retaining the original characteristics of the data. The final goal would be to find a latent space that would be the underlying parameters driving the evolution of galaxies. 
    
    \begin{figure}
        \centering
        \includegraphics[width=\linewidth]{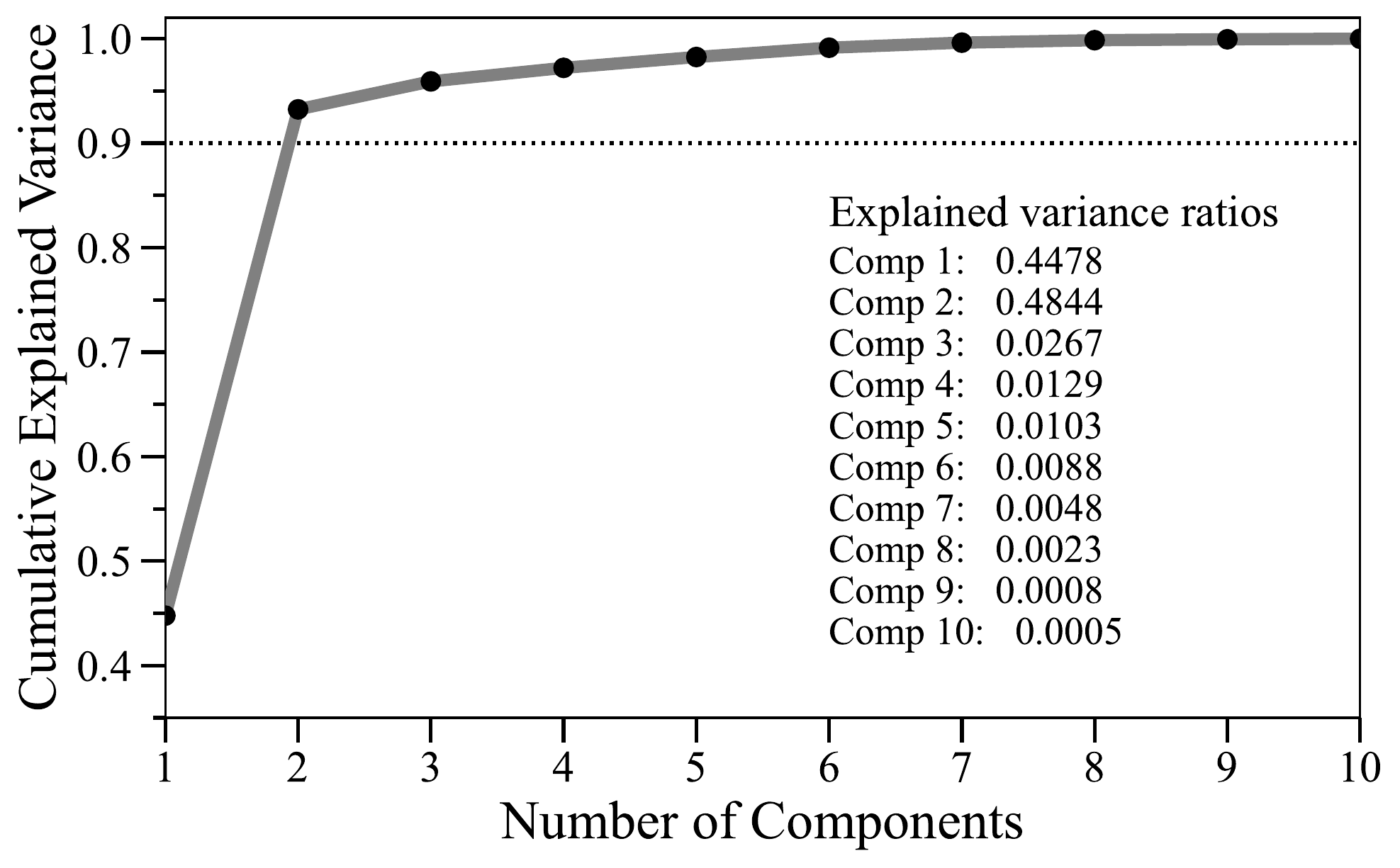}
        \caption{The cumulative explained variance of the data by the increasing number of manifold components. The first two components explain over 93\% of the variance in the multi-dimensional luminosity space for the sample.}
        \label{fig:cum_variance_explained}
    \end{figure}
        
    \begin{figure}
        \centering
        \includegraphics[width=\linewidth]{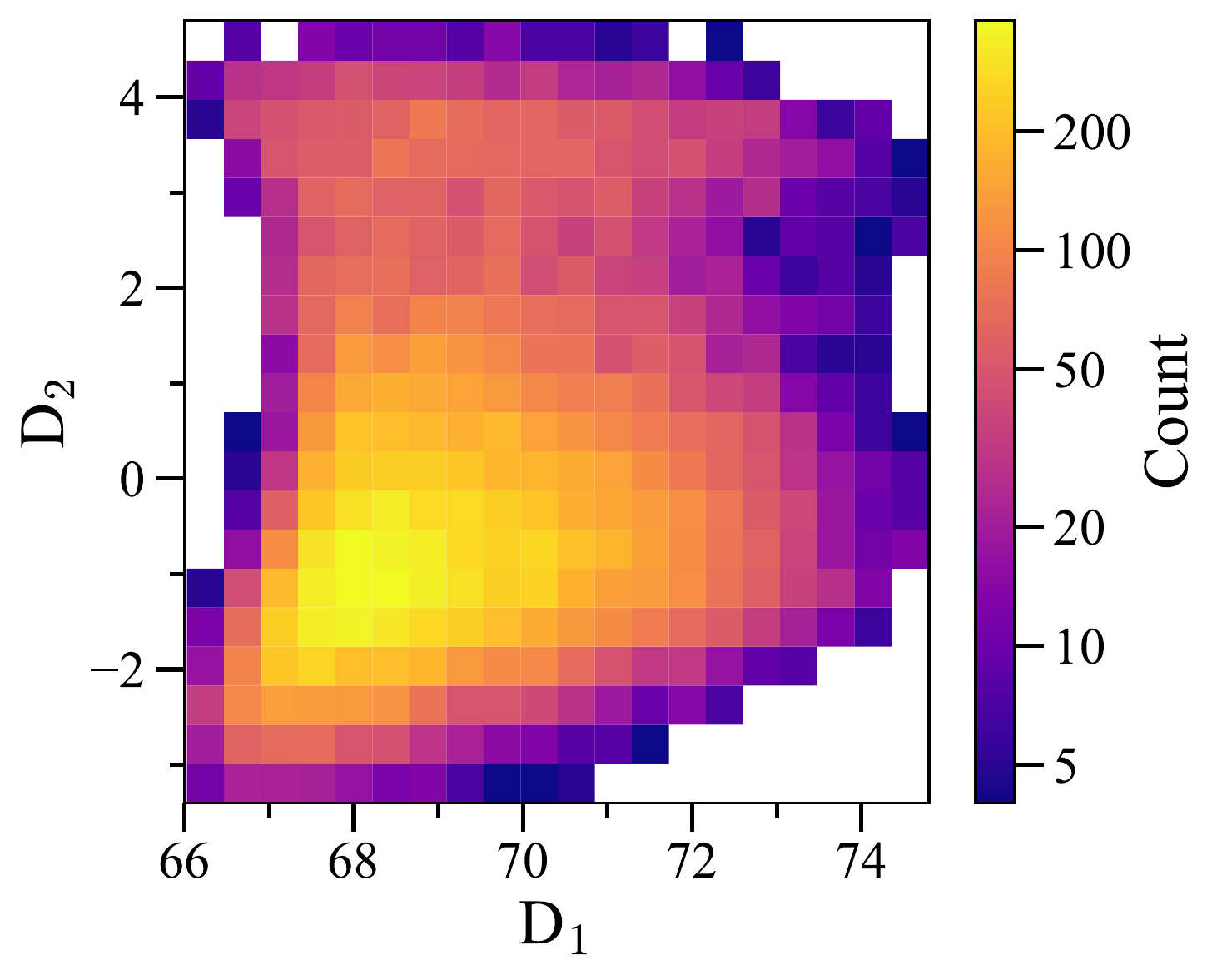}
        \caption{The distribution of galaxies on the manifold. A higher concentration of galaxies is found in the lower region left, which corresponds to the star-forming galaxies. The upper region corresponds to the quenched population and the region at around D$_2\approx$2 at the green-valley galaxies.}
        \label{fig:manifold_N}
    \end{figure}
    
    This study uses singular value decomposition (SVD) for dimensionality reduction. SVD is a matrix factorization technique that decomposes a matrix $A$ into three matrices in the form, 
    \begin{equation} \label{eq:svd}
        A = U \Sigma V^*.
    \end{equation}
    SVD generalizes the more commonly known eigendecomposition that acts on a normal square matrix to matrices with any shape $m \times n$. For a matrix $A$ shaped $m \times n$, $U$ is a unitary matrix of shape $m \times m$, $\Sigma$ is a $m \times n$ matrix with non-negative real numbers along the diagonal (also known as singular values), and $V$ is a $n \times n$ unitary matrix.
    
    A key application that allows SVD for dimensionality reduction is to estimate a low-rank approximation of the input matrix. Let us say that the low-rank approximation of the matrix $A$ is $\tilde{A}$ with rank($\tilde{A}$) $= r$. The SVD of $\tilde{A}$ is then given by,
    \begin{equation}\label{eq:trunc_svd}
        \tilde{A} = U \tilde{\Sigma} V^*,
    \end{equation}
    where $U$ and $V$ are the same as Eq. (\ref{eq:svd}), and $\tilde{\Sigma}$ is the same as $\Sigma$ with only the $r$ largest singular values as nonzero, while other smaller singular values are set to zero. The approximation is based on minimizing the Frobenius ($\ell_2$) norm of the difference in the reconstructed matrix and is called the Eckart–Young theorem. 
    
    SVD is often misunderstood as the principal component analysis (PCA) \citep[e.g.,][]{Conselice_2006}. The main difference is that PCA diagonalizes the covariance matrix, while SVD diagonalizes the data matrix. Though they qualitatively achieve similar results for dimensionality reduction, SVD produces analytic transformation matrices to move between the feature space (magnitudes) and the latent space, as provided later in this section.
    
    The data matrix with $\sim$27,000 galaxies (samples) of 11 magnitudes (features) are randomly split into train and test samples (70\% and 30\%, respectively) and fed into the \textsc{SKLEARN} implementation of SVD (\textsc{sklearn.TruncatedSVD}). Figure \ref{fig:cum_variance_explained} shows the dependence of cumulative variance ratio explained with each axis found by SVD. We find that the first and the second axes explain 44.78\% and 48.44\% variance of the data, which means that the data could be well approximated by two parameters with over 93.23\% explained variance. We call this two-dimensional structure the "Galaxy Manifold". 
    
    Figure \ref{fig:manifold_N} shows the number distribution of galaxies on the manifold. The region between D$_2\approx$ -2 and D$_2\approx$ 0 is the most densely populated. As we show later in Section \ref{sec:physical_properties}, the region corresponds to the star-forming blue cloud of galaxies. At around D$_2\approx$ 4, we have the passively star-forming population that can also be considered the quiescent population. We then have that D$_2\approx$ 2 corresponds to the transitional green-valley population.
    
\begin{figure*}
        \centering
        \includegraphics[width=\linewidth]{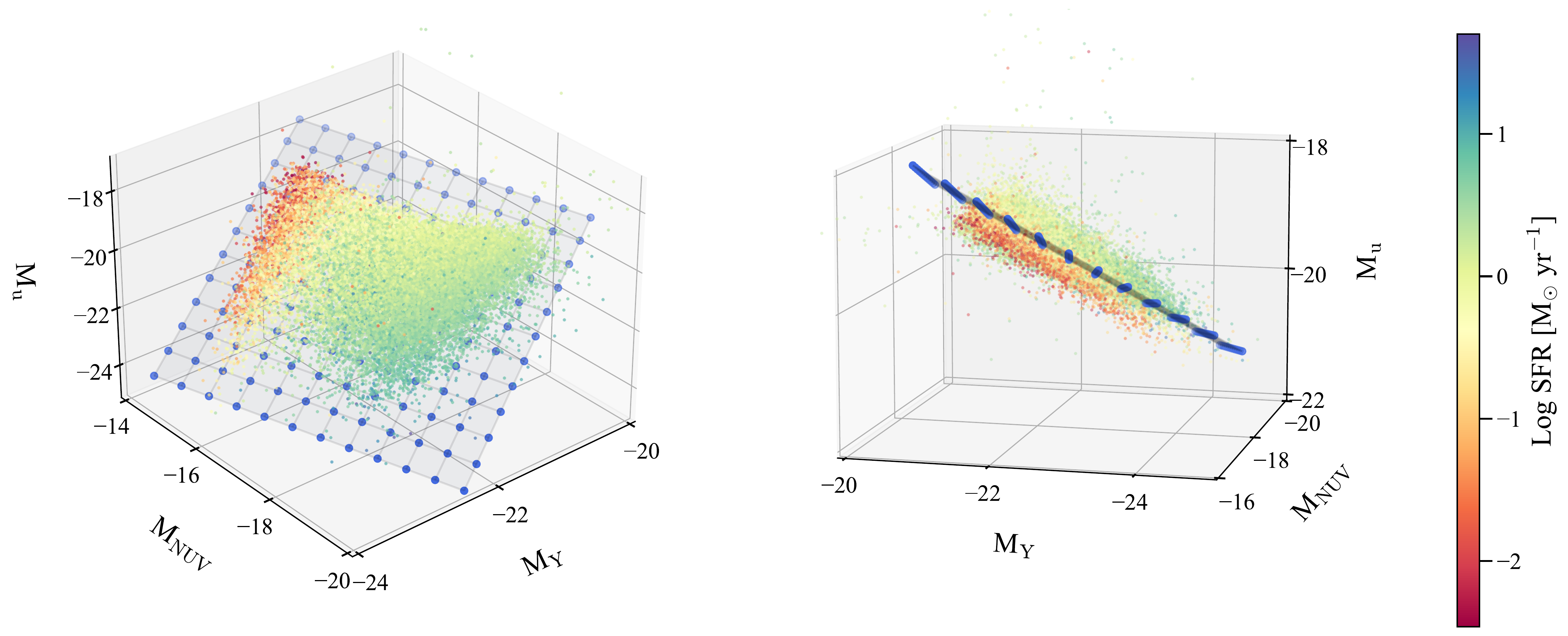}
        \caption{The found galaxy manifold from the 11-dimensional luminosity space shown within the 3D projected space of $M_{NUV}$, $M_{u}$, and $M_{Y}$. The left panel shows the edge on, and the right panel shows the edge on view of the manifold. Each point corresponding to each galaxy in the sample is colored based on its SFR, showing how the physical properties of galaxies change with manifold position.  }
        \label{fig:proj_headon_edgeon}
    \end{figure*}

    Figure \ref{fig:proj_headon_edgeon} shows the 3 dimensional projection of the 11-dimensional space in optical (\textsl{u}), ultraviolet (\textsl{NUV}), infrared (\textsl{Y}) with the SVD determined Galaxy Manifold shown head-on (left left) and edge-on (right panel). The colors of the dots correspond to the SFR derived in \citet{Salim_2016, Salim_2018}. We observe that galaxies are distributed along the discovered manifold and that key physical properties like SFR may be linked to the parameters of this manifold, which will be explored in the next section.
    
    Since we use a linear transformation technique to obtain the above manifold, the transformation can be written as a matrix multiplication to the data matrix. Provided the 11 magnitudes for a galaxy, one can use the transformation matrix to obtain the two parameters on the manifold. The forward transform is given by,
    \begin{equation} \label{eq:trans_matrix_11to2}
        \left[\begin{array}{c}
            \mathrm{D}_1^{\prime} \\
            \mathrm{D}_2^{\prime} \\
        \end{array}\right]
        = 
        \left[\begin{array}{cc}
             -0.2492 &   0.7078 \\
             -0.2594 &   0.5409 \\
             -0.2790 &   0.1340 \\
             -0.2981 &  -0.0076 \\
             -0.3064 &  -0.0758 \\
             -0.3106 &  -0.1137 \\
             -0.3133 &  -0.1428 \\
             -0.3201 &  -0.1687 \\
             -0.3214 &  -0.1988 \\
             -0.3258 &  -0.1992 \\
             -0.3217 &  -0.2043 \\
        \end{array}\right]^{\top} \left[\begin{array}{l}
            M_{\textit{FUV}} \\
            M_{\textit{NUV}} \\
            M_{\textit{u}} \\
            M_{\textit{g}} \\
            M_{\textit{r}} \\
            M_{\textit{i}} \\
            M_{\textit{z}} \\
            M_{\textit{Y}} \\
            M_{\textit{J}} \\
            M_{\textit{H}} \\
            M_{\textit{K}} \\
        \end{array}\right],
    \end{equation}
    where $M_{\mathrm{x}}$ is the magnitude in band {x}, and $x^{\prime}$ and $y^{\prime}$ correspond to the values on the manifold for a particular galaxy. Similarly, the backward transform is given by,
    \begin{equation} \label{eq:inv_trans_matrix_2to11}
        \left[\begin{array}{c}
            M_{\textit{FUV}}^{\prime} \\
            M_{\textit{NUV}}^{\prime} \\
            M_{\textit{u}}^{\prime} \\
            M_{\textit{g}}^{\prime} \\
            M_{\textit{r}}^{\prime} \\
            M_{\textit{i}}^{\prime} \\
            M_{\textit{z}}^{\prime} \\
            M_{\textit{Y}}^{\prime} \\
            M_{\textit{J}}^{\prime} \\
            M_{\textit{H}}^{\prime} \\
            M_{\textit{K}}^{\prime} \\
        \end{array}\right]
        = 
        \left[\begin{array}{cc}
             -0.2492 &   0.7078 \\
             -0.2594 &   0.5409 \\
             -0.2790 &   0.1340 \\
             -0.2981 &  -0.0076 \\
             -0.3064 &  -0.0758 \\
             -0.3106 &  -0.1137 \\
             -0.3133 &  -0.1428 \\
             -0.3201 &  -0.1687 \\
             -0.3214 &  -0.1988 \\
             -0.3258 &  -0.1992 \\
             -0.3217 &  -0.2043 \\
        \end{array}\right] \left[\begin{array}{l}
            \mathrm{D}_1^{\prime} \\
            \mathrm{D}_2^{\prime} \\
        \end{array}\right],
    \end{equation}
    where the $M'_x$ represent the approximate magnitude values derived from the two manifold values. The two-dimensional manifold can also be defined within the 11-dimensional space by the plane normal to the manifold with a set of 9 equations as Eq. (\ref{eq:manifold_eq}).
    \begin{figure*}
    \centering
    \normalsize
    \begin{equation}
    \resizebox{.9\hsize}{!}{$
        \label{eq:manifold_eq}
        \left[\begin{array}{rrrrrrrrrrr}
             -0.3863 &  0.0550 &  0.8103 &  0.1524 &  0.0988 &  0.0454 &  0.0038 & -0.0621 & -0.2754 & -0.1882 & -0.2025\\
              0.0063 & -0.0196 &  0.2658 & -0.1674 & -0.1844 & -0.1809 & -0.1574 & -0.0528 &  0.8478 & -0.0657 & -0.2890\\
              0.3100 & -0.4047 &  0.4117 & -0.2984 & -0.2520 & -0.1887 & -0.1066 & -0.1333 & -0.1276 &  0.2790 &  0.5091\\
             -0.4392 &  0.6566 & -0.0493 & -0.1160 & -0.1739 & -0.1653 & -0.1564 & -0.2453 &  0.0497 &  0.2226 &  0.4074\\
             -0.0636 &  0.1649 &  0.0548 & -0.3187 & -0.2769 & -0.1887 &  0.0055 &  0.8518 & -0.1457 & -0.0368 & -0.0604\\
              0.0066 & -0.0189 & -0.0086 &  0.0432 &  0.0270 &  0.0308 & -0.0006 &  0.0648 &  0.1483 & -0.8219 &  0.5425\\
             -0.0163 &  0.0972 &  0.0263 & -0.5699 & -0.0924 &  0.2341 &  0.7302 & -0.2282 & -0.0023 & -0.0989 & -0.0804\\
             -0.0030 &  0.0346 &  0.0228 & -0.4408 &  0.1824 &  0.6944 & -0.5354 &  0.0425 & -0.0036 &  0.0072 & -0.0056\\
             -0.0015 &  0.0166 &  0.0034 & -0.3673 &  0.8024 & -0.4685 & -0.0293 &  0.0182 &  0.0033 &  0.0028 &  0.0176
        \end{array}\right]%^{\top}
        \left[\begin{array}{l}
            M_{\textit{FUV}} \\
            M_{\textit{NUV}} \\
            M_{\textit{u}} \\
            M_{\textit{g}} \\
            M_{\textit{r}} \\
            M_{\textit{i}} \\
            M_{\textit{z}} \\
            M_{\textit{Y}} \\
            M_{\textit{J}} \\
            M_{\textit{H}} \\
            M_{\textit{K}} \\
        \end{array}\right]
         = 
        \left[\begin{array}{l}
            0 \\
            0 \\
            0 \\
            0 \\
            0 \\
            0 \\
            0 \\
            0 \\
            0 \\
        \end{array}\right]
        $}
    \end{equation}
    \end{figure*}

\section{Connection to the Physical Properties} \label{sec:physical_properties}

    This section identifies how galaxy physical properties vary on the manifold. The simple reason is that galaxies with varying physical properties should have different characteristics in the luminosity space. Additionally, since the galaxies distribute in a 2D plane within the luminosity space of far UV to near IR, the physical properties that can be estimated with those bands should also be best explained by the two dimensions. We show that various physical properties can be well represented on the manifold. For simplicity, we assume the physical properties used for the manifold calibration have negligible uncertainty.

\subsection{Stellar masses, star formation rates, and specific star formation rates}
\begin{figure*}
        \centering
        \includegraphics[width=0.8\linewidth]{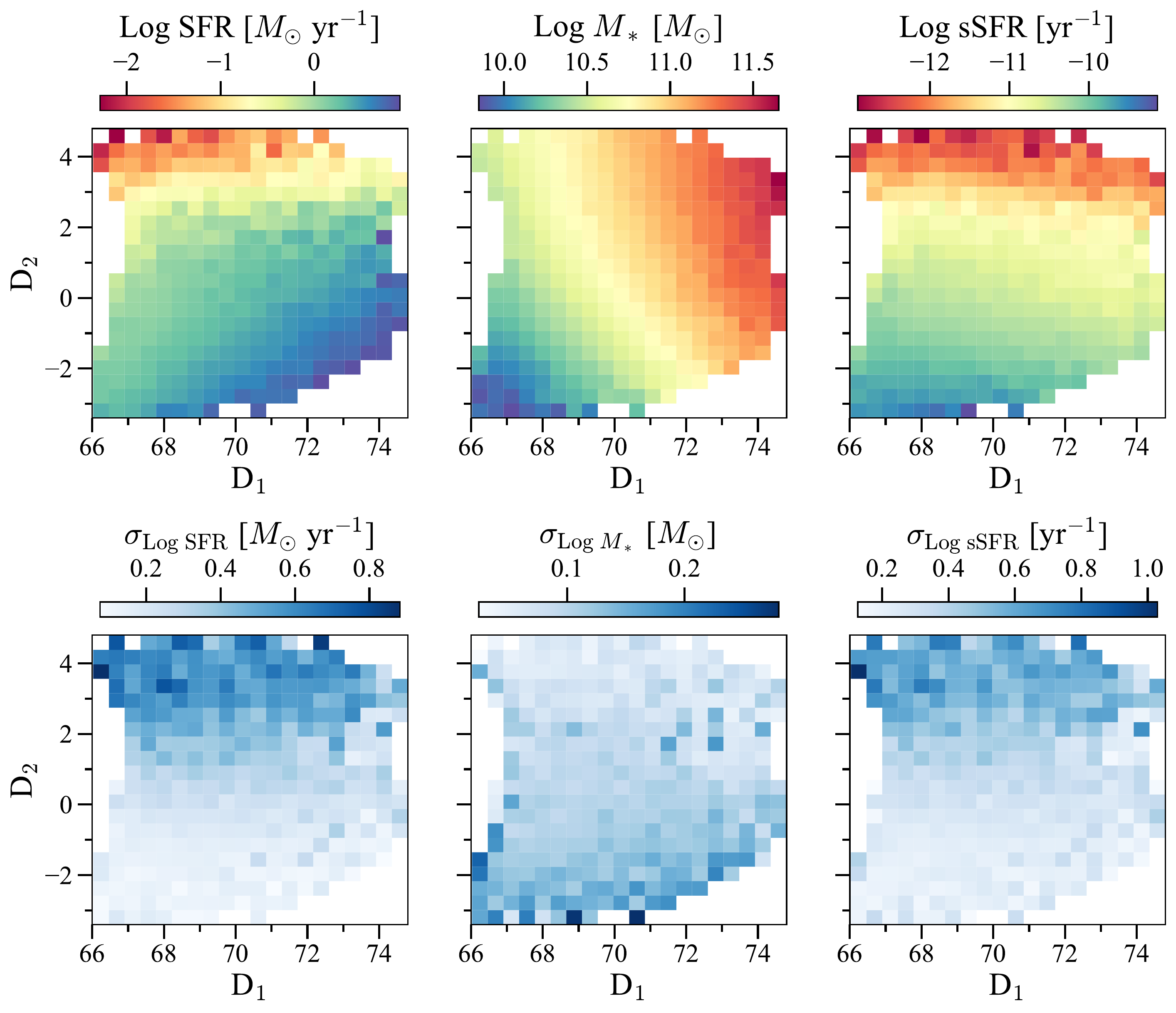}
        \caption{The distribution of star formation rates (SFR), stellar masses ($M_*$), and specific star formation rates (sSFR) on the galaxy manifold from left to right. For each property, the upper row shows the median values at each bin, and the lower row shows the standard deviation $\sigma$. The gradient of sSFR is closely related to D$_2$. SFR tends to have a higher scatter for lower SFR due to the difficulty of measurement, whereas $M_*$ has a higher scatter for less massive galaxies.}
        \label{fig:manifold_SF}
    \end{figure*}

    There is an elementary connection between luminosities and stars. Therefore, we examine the median $M_*$ and SFR values on the manifold. The $M_*$ and SFR values are derived in GALEX-SDSS-WISE LEGACY CATALOG \citep[GSWLC;][]{Salim_2016,Salim_2018}, which uses the SED fitting code CIGALE \citep{Burgarella_2005, Noll_2009, Boquien_2019}. Figure \ref{fig:manifold_SF} shows the distribution of SFR, $M_*$, and specific star formation rates (sSFR = SFR/$M_*$) on the galaxy manifold. The properties are binned and plotted on the manifold as color in the log scale. SFR decreases continuously from the bottom right towards the top left. The bluer regions roughly correspond to more actively star-forming galaxies, while redder regions correspond to the more quiescent galaxies. Lower SFR regions at the top appear to have a large scatter due to the difficulty in measurements. For $M_*$, we see a continuous change from lower-mass galaxies to massive galaxies. In the $M_*$ distribution, we see the increase from bottom-left to top-right and a smoother change along the manifold, with a smaller scatter seen for higher $M_*$ values. Interestingly, the D$_2$ is highly correlated with the sSFR of galaxies, where sSFR decreases when going up along the D$_2$. We interpret the above result that D$_2$ traces the evolutionary stage of galaxy star formation. The median values of $\sigma_{\mathrm{Log \ SFR}}$, $\sigma_{\mathrm{Log \ } M_*}$, and $\sigma_{\mathrm{Log \ sSFR}}$ are 0.34, 0.11, and 0.37, respectively. Surprisingly, the main sequence by definition in \citet{Renzini_2015} lies almost parallel to the $D_1$ at around D$_2\approx$-1.8.
    
\subsection{Gas mass}
\begin{figure}
        \centering
        \includegraphics[width=\linewidth]{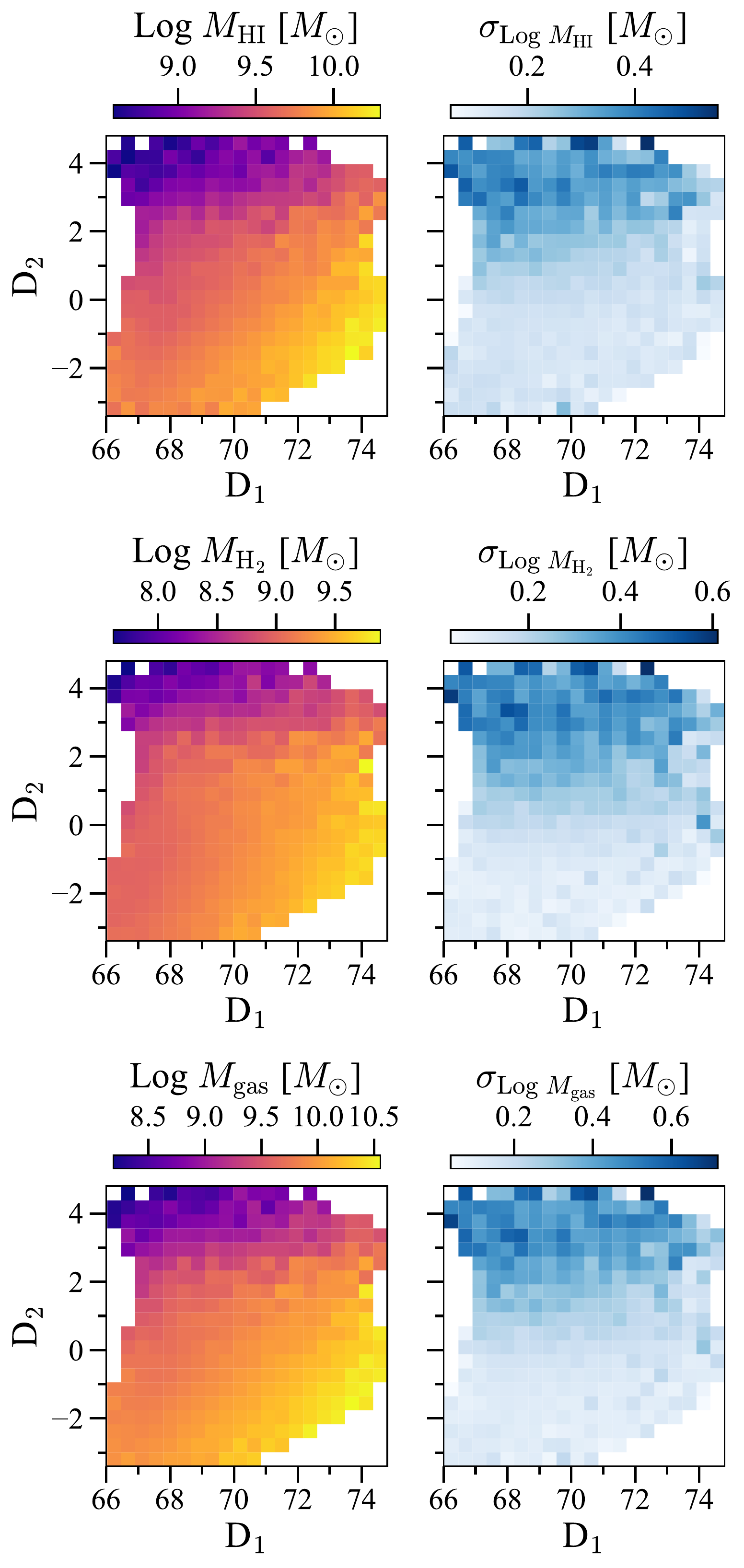}
        \caption{The distribution of masses of HI ($M_{\mathrm{HI}}$), H$_2$ ($M_{\mathrm{H_2}}$) and total gas ($M_{\mathrm{gas}}$) derived from \citet{Yesuf_2019} shown on the galaxy manifold. For each property, the left column shows the median values at each bin, and the right column shows the standard deviation $\sigma$. Generally, all three gas mass values decrease from the lower right to the upper left, and lower mass regions tend to have higher scatter.}
        \label{fig:manifold_gas}
    \end{figure}

    A key property that controls the star formation of a galaxy is gas. However, observations of neutral atomic hydrogen (HI) and molecular hydrogen (H$_2$) are demanding, limiting the number of galaxies with observed HI and H$_2$ masses. The galaxies with these measurements also tend to be gas-rich, which may bias the distribution on the manifold. Thus, we resort to using empirical relations provided in \citet{Yesuf_2019} in deriving gas mass estimates using dust absorption, galaxy size, $M_*$, and SFR of a galaxy. The above work also provides the total gas mass, which will be used for discussion in Section \ref{sec:discussion}. The equations used to derive the masses for HI ($M_{\mathrm{HI}}$), H$_2$ ($M_{\mathrm{H_2}}$), and total gas masses ($M_{\mathrm{gas}}$) are given below. 
    \begin{equation}\label{eq:HI}
        \begin{split}
            \log M_{\mathrm{HI}}=(9.07\pm0.04) + (1.08&\pm0.11) \log R_{50} \\
                            &+ (0.47\pm0.02) \log \operatorname{SFR},
        \end{split}
    \end{equation}
    \begin{equation}\label{eq:H2}
        \begin{split}
            \log M_{\mathrm{H}_{2}} &= (6.56\pm0.37) + (0.41\pm0.09) A_{V} \\ &+ (0.30\pm0.10) \log R_{50} + (0.21\pm0.04) \log M_*\\ &+(0.61\pm0.03) \log \operatorname{SFR},
        \end{split}
    \end{equation}
    \begin{equation}\label{eq:Tgas}
        \begin{split}
            \log M_{\mathrm{gas}}=(9.28\pm0.04) + (0.87&\pm0.11) \log R_{50} \\
                            &+ (0.70\pm0.04) \log \operatorname{SFR},
        \end{split}
    \end{equation}
    where $R_{50}$ is the half-light radius of the galaxy, and $A_V$ is the dust attenuation at V-band obtained from the RCSED catalog. The derived quantities using the above relations are shown in Figure \ref{fig:manifold_gas}. All three plots show a gradual increase from left bottom to top left. The median values of $\sigma_{\mathrm{Log \ } M_{\mathrm{HI}}}$, $\sigma_{\mathrm{Log \ } M_{\mathrm{H_2}}}$, and $\sigma_{\mathrm{Log \ } M_{\mathrm{gas}}}$ are 0.21, 0.24, and 0.26, respectively.

\subsection{Galaxy morphology} \label{sec:physical_properties_morph}
\begin{figure}
        \centering
        \includegraphics[width=\linewidth]{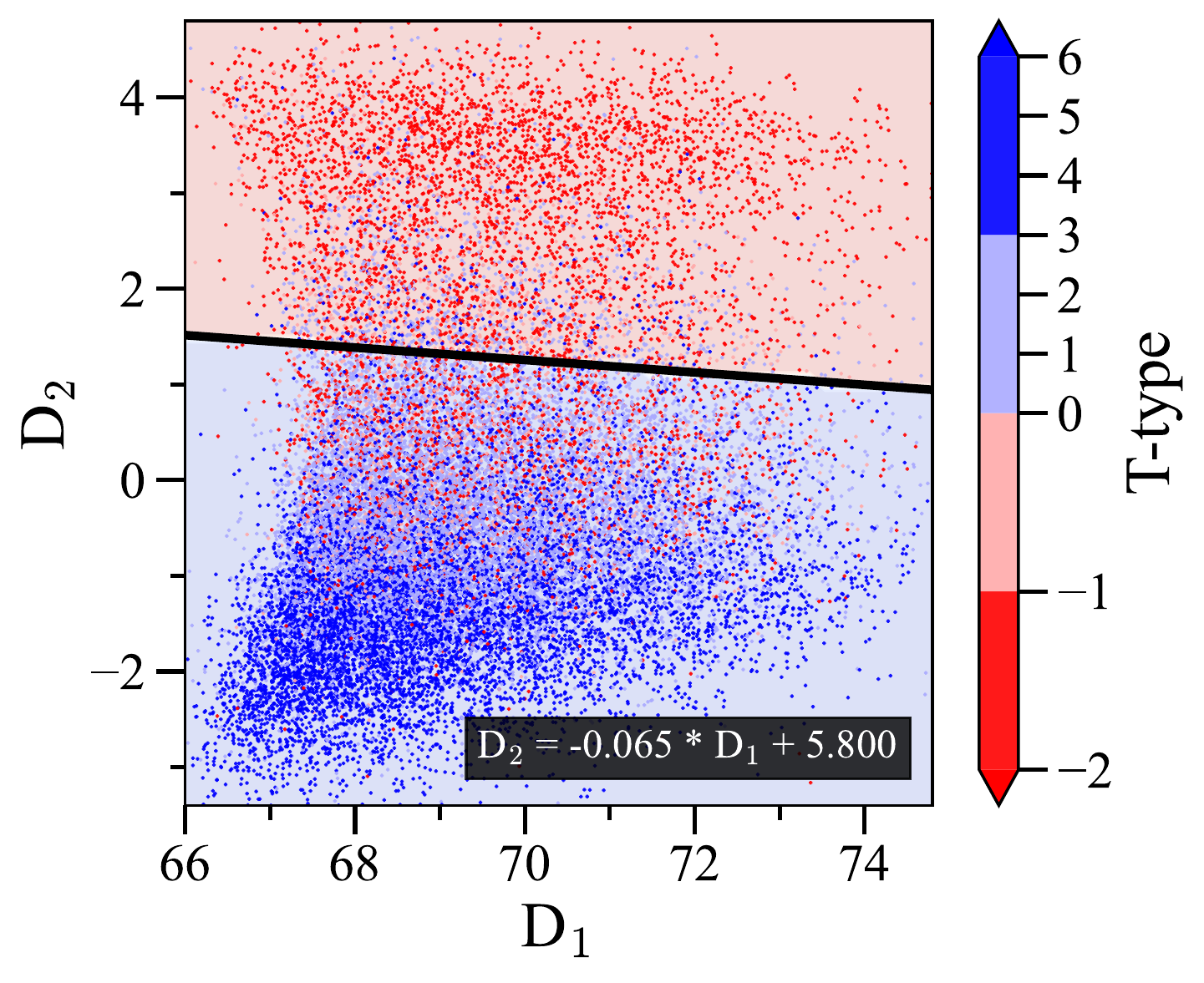}
        \caption{The distribution of galaxies on the manifold colored based on the Hubble T-type derived in \citet{Dominguez_2018}. Blue/red points correspond to spiral/elliptical galaxies. The black line shows the morphology type boundary determined by logistic regression. }
        \label{fig:manifold_Ttype}
    \end{figure}

    We look at the morphologies provided in  \citet{Dominguez_2018}, which includes the Hubble T-types and Galaxy Zoo 2 classifications using convolutional neural networks. Both color and morphology are significantly related, with most blue galaxies being late-types (spirals) and most early-types (ellipticals) being red. Color is often used as a handy selection criterion for morphological classification \citep[e.g.,][]{Smethurst_2022}. This section tries to understand the relationship between the manifold axes and morphology. Figure \ref{fig:manifold_Ttype} shows the distribution of T-types along the manifold. Early-type galaxies correspond to T-type>0, spirals (Sa to Sm) are T-type<0, and T-type=0 are S0 galaxies. There is a clear separation of T-types in the D$_1$ and D$_2$ space, which suggests that the manifold axes can be a criterion for morphological classification. 
    
    The classification was done with logistic regression, where we considered T-type>0 and T-type$\le$0 as two labels. The obtained boundary was obtained to be,
    \begin{equation}
        \mathrm{D}_2 =-0.065*\mathrm{D}_1 + 5.800,
    \end{equation}
    where we obtained an accuracy of 0.85, with 1.0 as the best classification. We also found that adding more manifold axes to the regression did not improve the classification accuracy, implying that two dimensional manifold already provides sufficient information for classification.
    
\begin{figure}
        \centering
        \includegraphics[width=\linewidth]{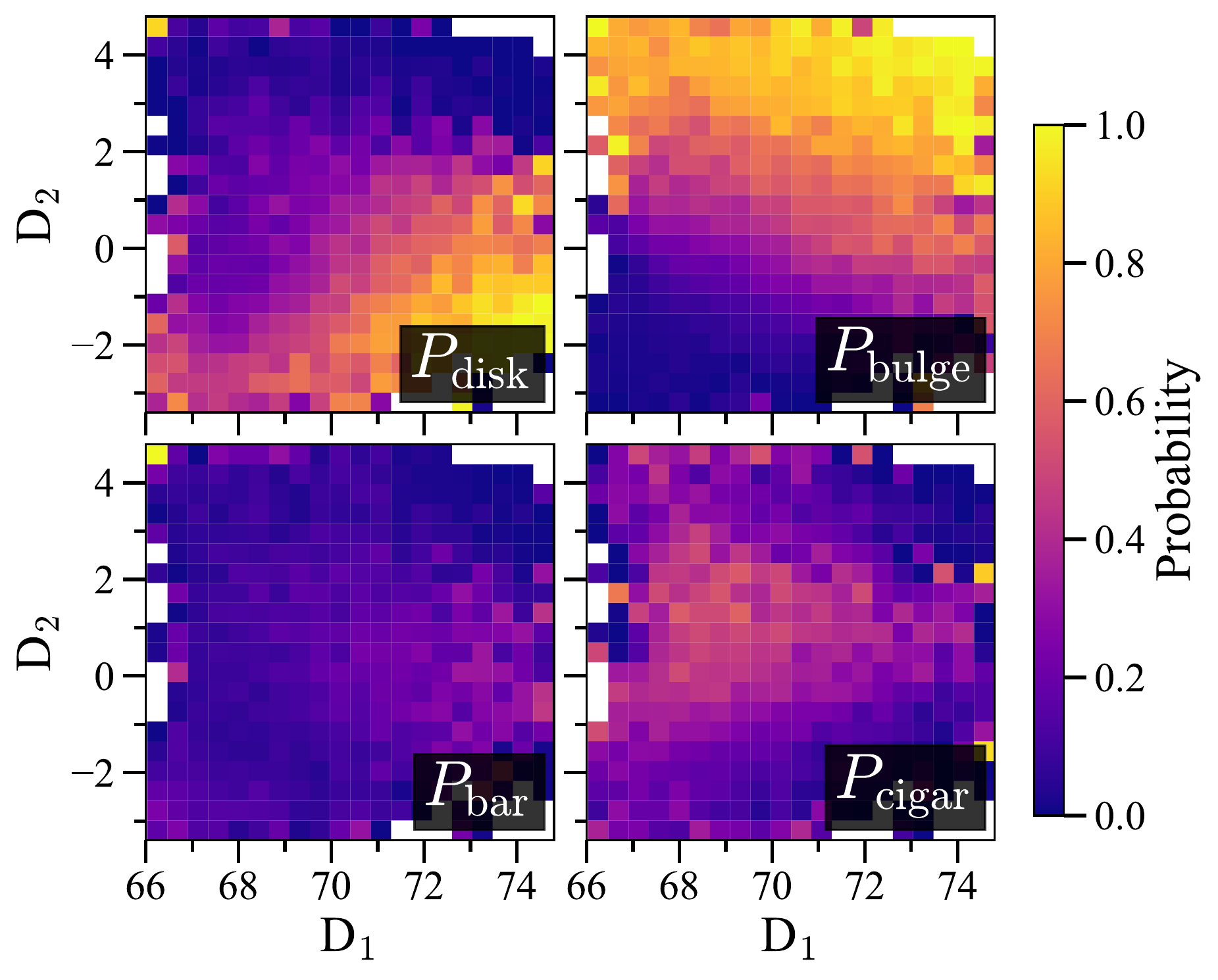}
        \caption{The probability distributions on the manifold of finding a disk (top left panel), bulge (top right panel), bar (bottom left panel), and cigar (bottom right panel) features in a galaxy. Brighter color corresponds to a higher median probability of a galaxy with that feature. Probabilities of morphological properties are derived from \citet{Dominguez_2018}.}
        \label{fig:manifold_morph}
    \end{figure}

    The distribution of median probabilities of Galaxy Zoo morphologies are shown in Figure \ref{fig:manifold_morph}. The concerned morphological features are disk (top left panel), bulge (top right panel), bar (bottom left panel), and cigar (bottom right panel). We see the trivial relationships between morphology and the location, such as elliptical having more bulges and star-forming galaxies more often with disks. However, more surprisingly, cigars tend to be located more often in the green valley.

\section{Evolution on the Manifold} \label{sec:evolution}

    If the currently observed galaxies exist on the manifold, we can deduce that galaxies should also evolve on the manifold. When galaxies evolve, their physical properties also change. Thus, we can express the evolution as a function of the two manifold parameters. If current galaxies are well-expressed on the manifold, their evolution should also be well-represented on this manifold. 

\subsection{Constant star formation evolution} \label{sec:Evolution-constSFR}

    The simplest evolution scenario is to assume a constant SFR for a given period $\Delta t$. We could think of a situation with infalling gas to sustain ongoing star formation. This mode of evolution is similar to a situation in the star-forming stage of a galaxy \citep[e.g.,][]{Bouche_2010, Daddi_2010, Genzel_2010, Tacconi_2010, Dave_2012, Dekel_2013, Lilly_2013,Forbes_2014a, Forbes_2014b, Hopkins_2014,Mitra_2015,Tacchella_2016}. If a galaxy starts at a particular location $[d1(t_1), d2(t_1)]$ on the manifold at time $t_1$, we can define SFR($t_1$) and $M_*$($t_1$) as a function of $d1(t_1)$ and $d2(t_1)$ as input (c.f. Appendix \ref{sec:fitting}). Thereafter, at time $t_2 = t_1 + \Delta t$, we have that $M_*(t_2) = M_*(t_1)+ (1-r) \cdot \operatorname{SFR}(t_1)\Delta t$, where $r=0.35$ is the return fraction assuming a Chabrier IMF \citep{Chabrier_2003}. We set that $\operatorname{SFR}(t_2) = \operatorname{SFR}(t_1)$. Since the SFR and $M_*$ maps have almost orthogonal gradients, choosing SFR and $M_*$ also gives us the values on the manifold.
    
\begin{figure}
        \centering
        \includegraphics[width=\linewidth]{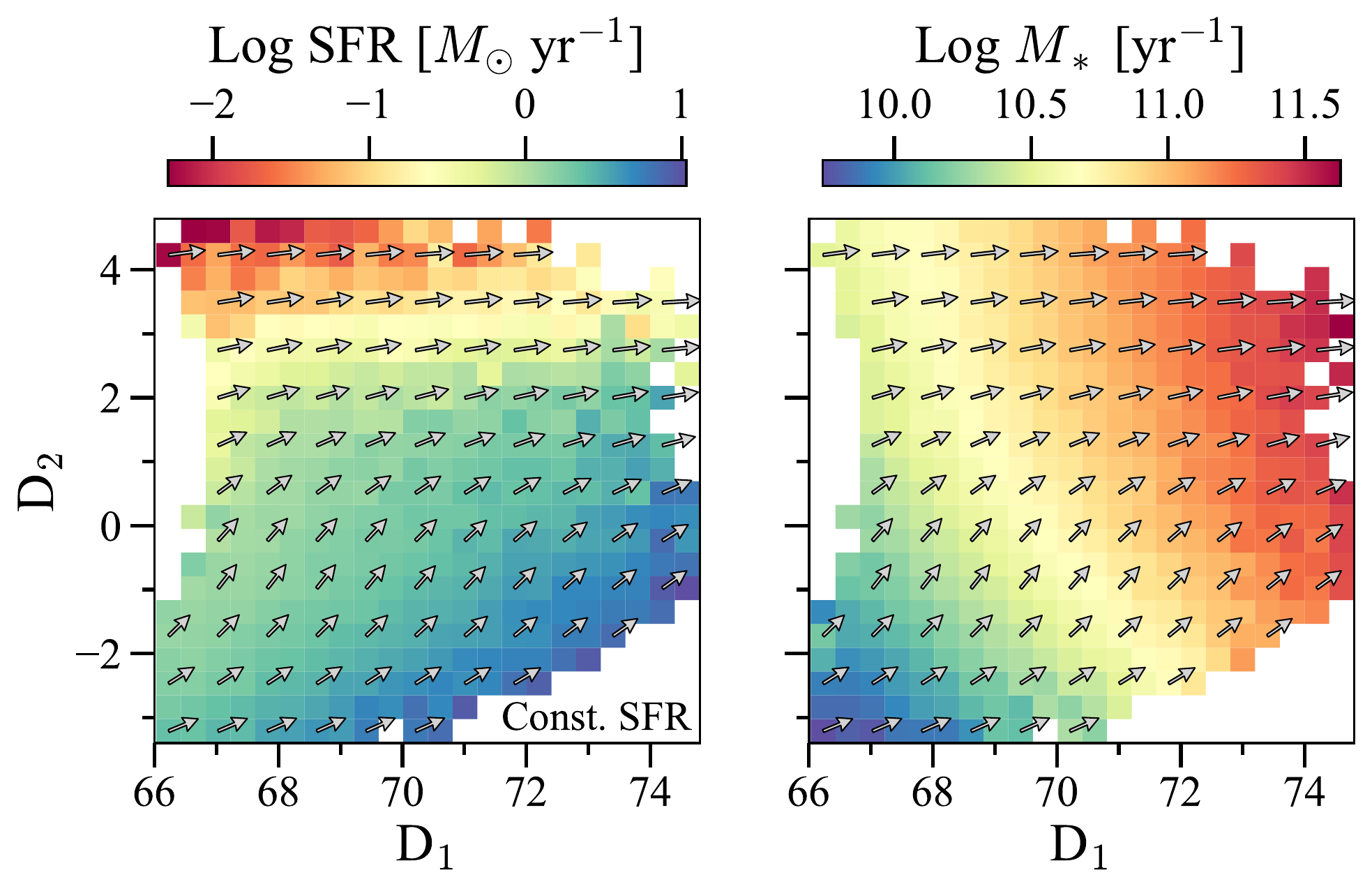}
        \caption{The star formation rate (left panel) and stellar mass distributions (right panel) over-plotted with arrows showing the direction of evolution on the manifold with constant star formation rate. Arrows generally point leftward along the "iso-SFR" lines towards increasing stellar masses. The directions are also perpendicular to the "iso-$M_*$," suggesting that this is the most efficient mode of increasing stellar mass in the galaxy.}
        \label{fig:evo-constant-norm}
    \end{figure}

    By considering galaxies sampled on a grid, we derive a vector field for the manifold. We calculate the direction in which a galaxy will move on the manifold given constant SFR. Figure \ref{fig:evo-constant-norm} shows the "vector field" on SFR and $M_*$ maps. The arrow lengths are normalized to a constant for visibility. In reality, the largest change in amplitude occurs for the most actively star-forming less massive galaxies, which occupy the lower regions. The movement on the manifold per unit of time is smaller for quiescent galaxies that are more massive and make fewer stars. Arrows follow the direction of increasing $M_*$ along the "iso-SFR" lines. Since the arrows point along the gradient of the $M_*$ map, it can be understood that constant SFR is an efficient mode of evolution to grow the $M_*$ of galaxies. If all the galaxies on the manifold receive enough gas to sustain the star formation for an extended period, the manifold should move along the arrow direction. While this may apply to the main sequence galaxies, this may not be plausible for galaxies that have left the main sequence.
    
    % \begin{figure*}
    %     \centering
    %     \includegraphics[width=0.45\linewidth]{manifold_sfr_evo_constant.pdf}
    %     \includegraphics[width=0.45\linewidth]{manifold_sm_evo_constant.pdf}
    %     \caption{Caption}
    %     \label{fig:manifold_evo-constant}
    % \end{figure*}
    
\subsection{Exponentially declining star formation evolution} \label{sec:Evolution-declSFR}

\begin{figure}
        \centering
        \includegraphics[width=\linewidth]{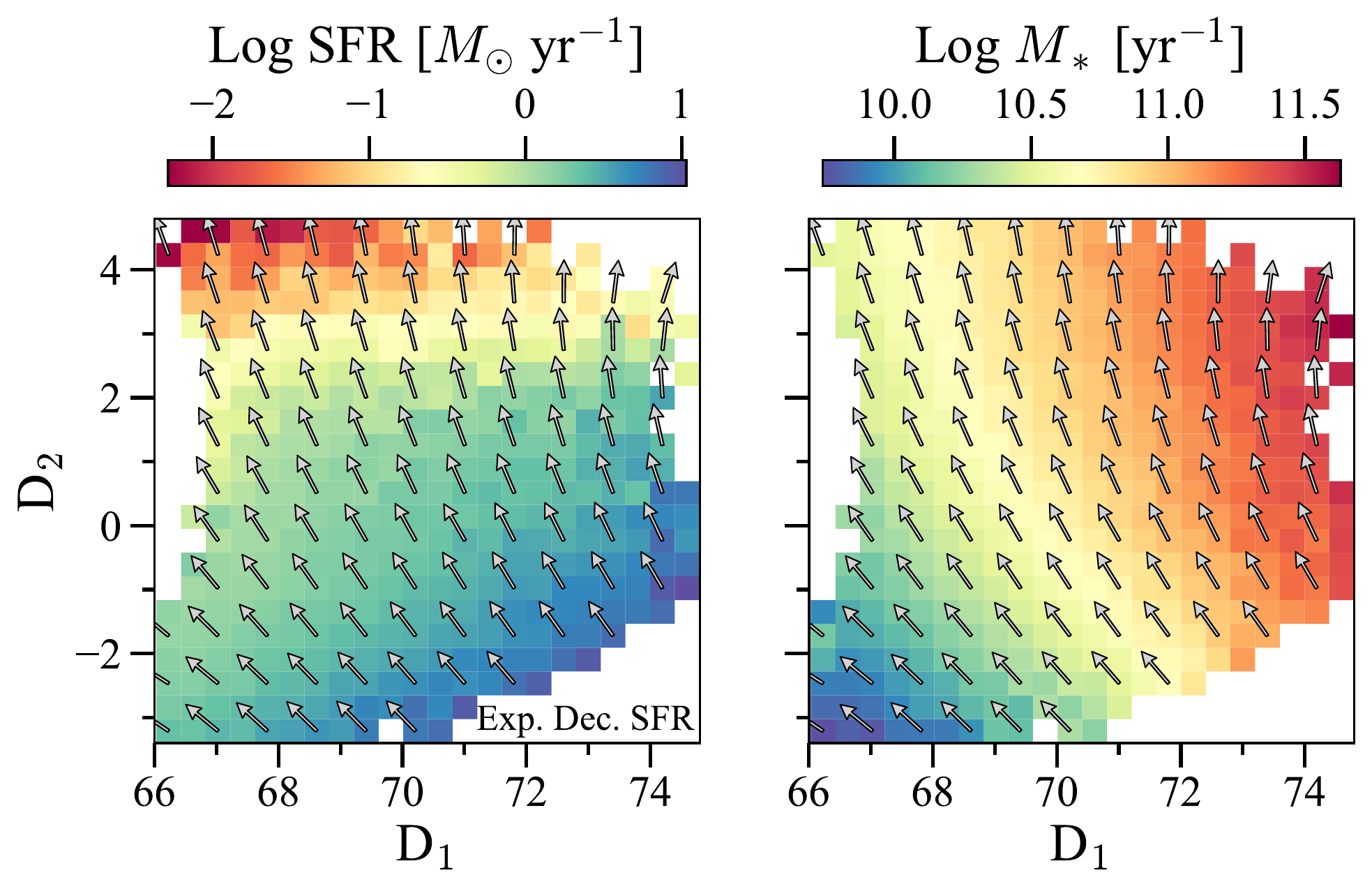}
        \caption{The star formation rate (left panel) and stellar mass distributions (right panel) over-plotted with arrows showing the direction of evolution on the manifold with exponentially declining star formation rate (decay timescale of 1 Gyr). Arrows point towards lower star formation rates (upward) and almost along the "iso-$M_*$" with a slight increase in stellar masses every timestep.}
        \label{fig:evo-exponential-norm}
    \end{figure}

    Here we consider the case where star formation declines exponentially. It can be understood as the scenario where there is no infall of gas in a galaxy, and the star formation declines exponentially, consuming the existing gas within the galaxy \citep[e.g.,][]{Faber_2007,Peng_2010, Schaye_2010, Renzini_2016}. We consider SFR($t_2$) = SFR($t_1$) $\cdot$ exp(-$t/\tau$), where $\tau$ is the decay timescale that is assumed to be 1 Gyr \citep{Young_1991,Daddi_2010,Genzel_2010,Saintonage_2011b,Tacconi_2013,Saintonage_2013}. Similarly to the above section, we calculate the SFR($t_2$) and $M_*$($t_2$) and derive the directions the galaxy moves as shown in Figure \ref{fig:evo-exponential-norm}. In the assumed model, arrows point toward SFR decline along "iso-$M_*$" lines with slowly increasing stellar masses.
    
    While the constant SFR galaxies move toward the upper right of the manifold, the exponentially declining SFR galaxies will move toward the upper left. Since the two vector fields corresponding to the two scenarios are mostly orthogonal, we extrapolate that real galaxies evolve on the manifold by a combination/episodes of the two evolutionary modes. 
    
\subsection{Evolution of gas-regulated systems}

    Assuming star formation histories often as analytic functions can be a strong assumption \citep[e.g.,][]{Carnall_2018}. 
    Thus, without assuming the star formation history, we consider a simple galaxy model like the ones of \citet[e.g.,][]{Tinsley_1980}, where $M_{\mathrm{gas}}$ and the $M_*$ evolve. Since we have both the $M_{\mathrm{gas}}$ and $M_*$ measurements, we remove the necessity of an SFH assumption.
    Here we consider the case where gas accretion has shut down and the two quantities $M_{\mathrm{gas}}$ and $M_*$ evolve as follows.
    \begin{equation}
        \begin{aligned}
            &M_{*}\left(t_{n+1}\right)=M_{*}\left(t_{n}\right)+(1-r) \cdot \operatorname{SFR}\left(t_{n}\right) \cdot d t \\
            &M_{\mathrm{gas}}\left(t_{n+1}\right)=M_{\mathrm{gas}}\left(t_{n}\right)-(1-r+\eta) \cdot \operatorname{SFR}\left(t_{n}\right) \cdot d t,
        \end{aligned}
    \end{equation}
    where $\eta$ is the "mass-loading" factor representing wind-driven gas mass loss. Observations give a wide range of values for $\eta$, ranging from 0.1 to 30 \citep[e.g.,][]{Bouche_2012,Newman_2012,Bolatto_2013,Kacprzak_2014,Schroetter_2015,Schroetter_2019, Davies_2019b,Forster-Schreiber_2019,Kruijssen_2019,Chevance_2020}. The value of $\eta$ appears to be weakly dependent on the redshift and $M_*$, which is also predicted in theoretical models \citep{Barai_2015,Muratov_2015,Torrey_2019}. For simplicity, $\eta$ is given a constant value of 2.5 as in \citet{Andrews_2017}. $\eta$ is essentially the free parameter that decides the trajectory on the manifold.
    
    $M_*$ evolve similarly to the one considered in section \ref{sec:Evolution-constSFR}. A key feature of this model is that SFR at a particular time is regulated by the gas mass \citep[gas-regulated;][]{Lilly_2013} present in some reservoir of the system through the star formation efficiency (SFE=SFR/$M_{\mathrm{gas}}$). On the manifold, we can know both the SFR and $M_{\mathrm{gas}}$ given the manifold axis values. Removing the need to assume a SFE is an added benefit of this technique. With this setup, it is also not necessary to consider a star formation history (SFH) because the next location on the manifold is decided completely on the above equations, which gives us manifold location, which provides us the SFR, $M_*$ and $M_{\mathrm{gas}}$ information necessary to calculate the next time step.
    
\begin{figure*}
        \centering
        \includegraphics[width=0.9\linewidth]{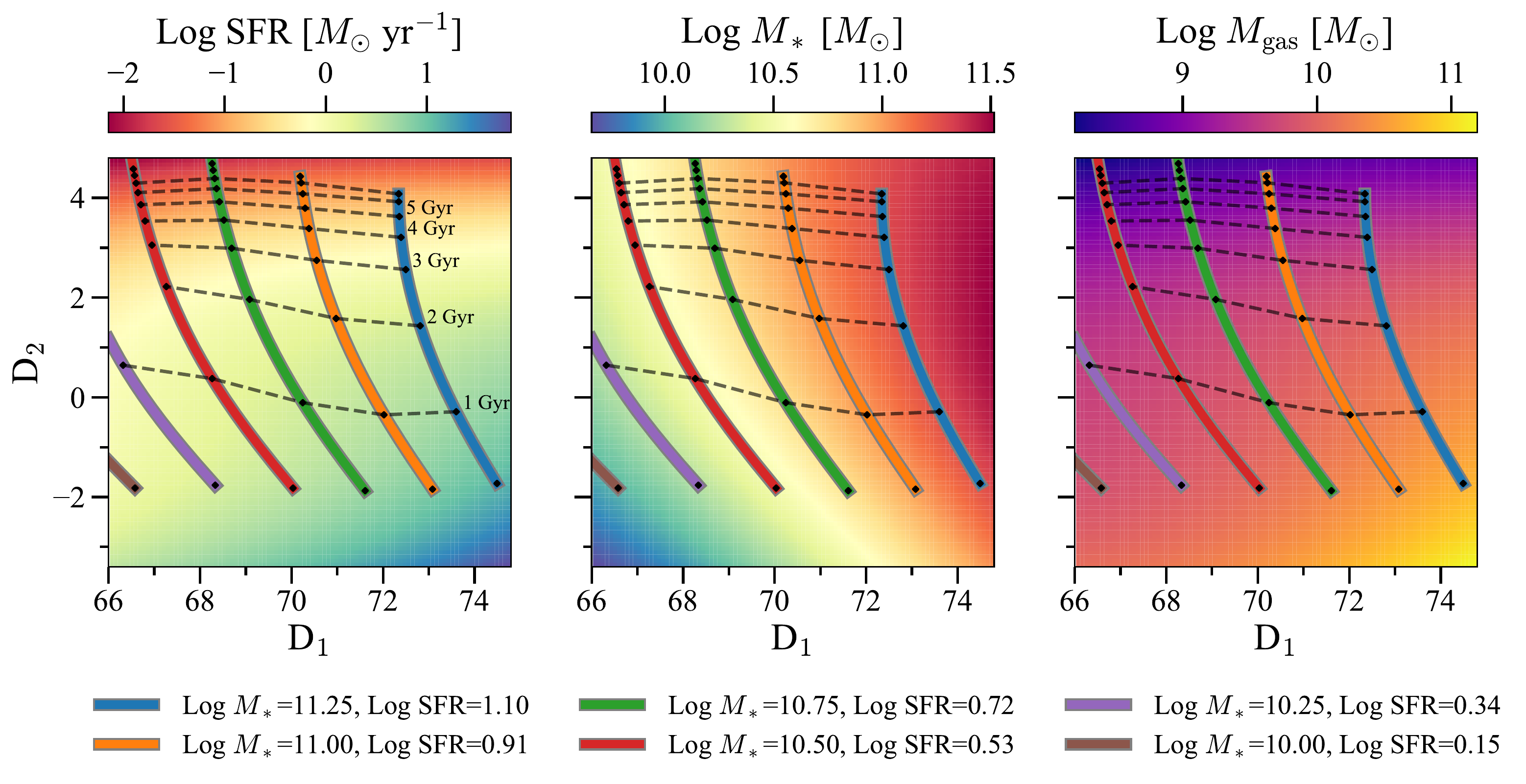}
        \caption{The evolution tracks on the SFR, $M_*$, and $M_{\mathrm{gas}}$ maps derived independently of an assumption of star formation history. The maps shown are the 2D cubic fit surfaces for each property, and the solid-colored thick lines correspond to each galaxy track}
        \label{fig:Evo-gasmodel}
    \end{figure*}

    Galaxies evolve along the star-forming sequence until a quenching event occurs \citep[e.g.,][]{Peng_2010}. Here we focus on how the galaxies evolve after leaving the main sequence. We consider galaxies of Log ($M_*$/$M_{\odot}$) in the range [10, 11.25] at 0.25 interval and calculate the evolution tracks according to the above procedure with $dt$=10 Myr. Figure \ref{fig:Evo-gasmodel} shows the evolution tracks on the SFR, $M_*$, and $M_{\mathrm{gas}}$ maps. The maps shown are the 2D cubic fit surfaces for each property, and the solid-colored thick lines correspond to each galaxy track. The tracks evolve towards the top left, and black dots joined with the thin black lines represent the location every 1 Gyr since leaving the main sequence (MS). Qualitatively the tracks also move along the vector field shown in Figure \ref{fig:evo-constant-norm}.
    
\begin{figure}
        \centering
        \includegraphics[width=0.9\linewidth]{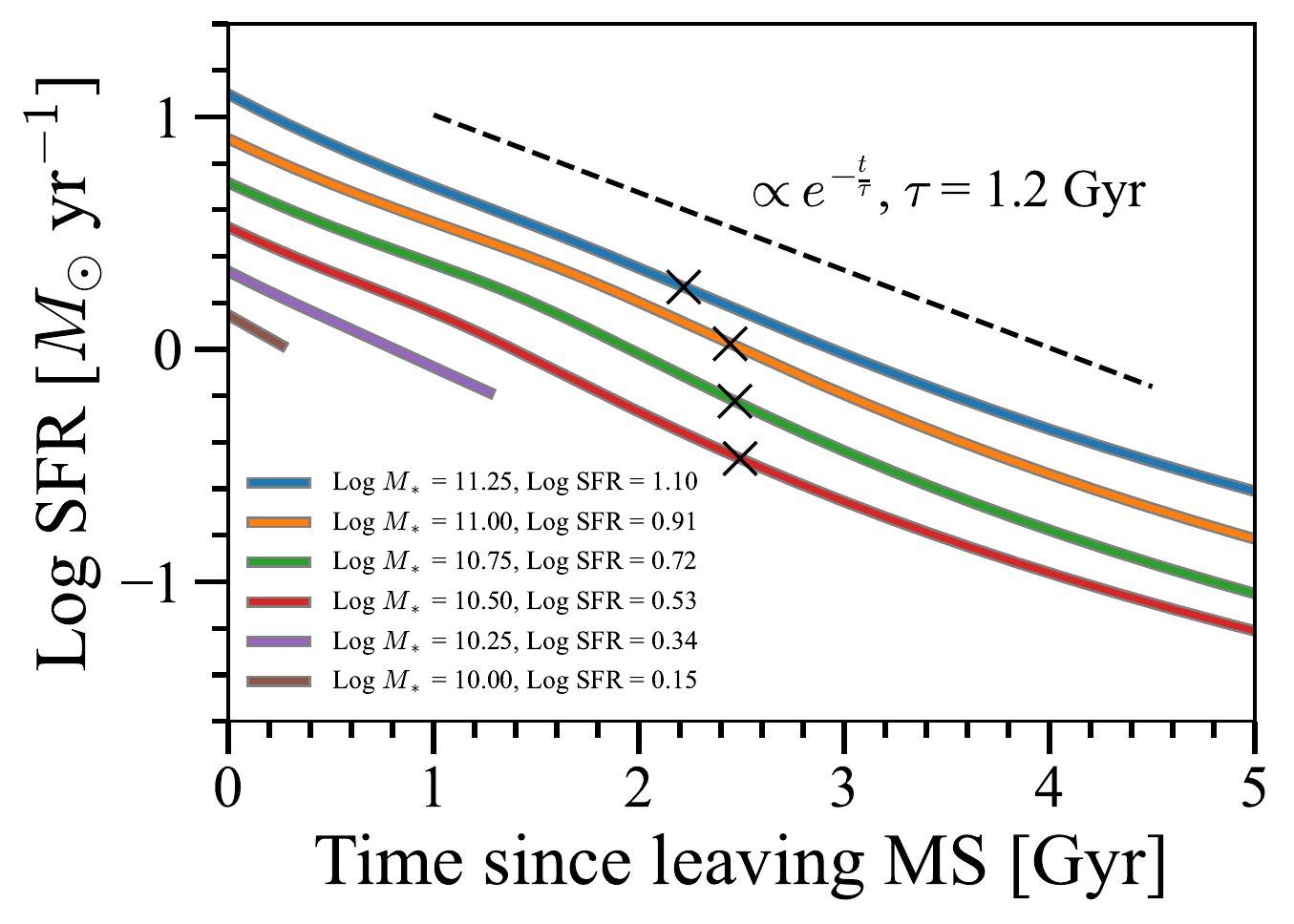}
        \caption{Star formation histories of the tracks shown in Figure \ref{fig:Evo-gasmodel}. SFRs decreases roughly exponentially, with a timescale of 1.2 Gyrs once it leaves the main sequence (MS). The crosses correspond to the time at which the galaxy quenches (Log sSFR = -11 [yr$^{-1}$]).}
        \label{fig:Evo-gasmodel-sfr}
    \end{figure}

    Figure \ref{fig:Evo-gasmodel-sfr} shows the derived SFHs for each of the tracks using the above procedure. All the galaxy tracks roughly follow an exponentially declining SFH with $\tau \sim 1.2$ Gyrs. The black crosses correspond to the time of quenching, which we consider the time the galaxies have Log sSFR $\le$ -11 [yr$^{-1}$]. This crossing time is approximately 2.5 Gyrs for all the tracks. However, there are some caveats to this model. Firstly, we assume the gas inflows to be zero. Galaxies continue their star formation within the MS through depletion and replenishment of gas \citep{Tacchella_2016}. Therefore, realistically, galaxies tend to be replenished with gas, with more significant and extended inflows expected for smaller-mass galaxies. Another effect is that we have assumed the same mass loading factor $\eta$. More realistically, $\eta$ should be given a mass dependence such as the relation provided in \citet{Muratov_2015}. We find that these effects the tracks in varying degrees, resulting in different slops for the SFH and crossing times. 
    
\section{Discussion} \label{sec:discussion}

The two axes of the Galaxy Manifold express over 93\% of the variance of our sample. Firstly, we discuss the remaining variance unexpressed by the two axes in Section \ref{sec:discussion_unexplained_variance}. Secondly, we discuss the physical interpretation of the two axes (Section \ref{sec:discussion_interpretation}) and the identified evolutionary modes (Section \ref{sec:discussion_2modes}). After that, in Section \ref{sec:discussion_property_predict}, we test the accuracy of estimating physical properties using the manifold. Section \ref{sec:discussion_recover} explores the idea of recovering the manifold axes without the full features (11 bands) used for the transformation matrix defined in Eq. (\ref{eq:trans_matrix_11to2}). We compare our analysis with the non-linear dimensionality reduction technique in Section \ref{sec:discussion_SOM}, and lastly, some limitations of this work are presented in Section \ref{sec:discussion_limitations}.

\subsection{Unexplained variance by the two axes}  \label{sec:discussion_unexplained_variance}

    As shown in Figure \ref{fig:cum_variance_explained}, we cannot represent all the information about galaxies in 2 dimensions. The trivial solution is considering more dimensions/components from the SVD analysis. For completeness, we provide the full transformation matrix in Equation \ref{eq:full_trans_matrix}, which allows one to obtain the 10 orthogonal components. 
    \begin{figure*}
    \centering
    \normalsize
    \begin{equation}\label{eq:full_trans_matrix}
    \resizebox{.9\hsize}{!}{$
        \left[\begin{array}{c}
            \mathrm{D}_1^{\prime} \\
            \mathrm{D}_2^{\prime} \\
            \mathrm{D}_3^{\prime} \\
            \mathrm{D}_4^{\prime} \\
            \mathrm{D}_5^{\prime} \\
            \mathrm{D}_6^{\prime} \\
            \mathrm{D}_7^{\prime} \\
            \mathrm{D}_8^{\prime} \\
            \mathrm{D}_9^{\prime} \\
            \mathrm{D}_{10}^{\prime} \\
        \end{array}\right]
        = 
        \left[\begin{array}{rrrrrrrrrrr} % = V*
            -0.2492 & -0.2594 & -0.2790 & -0.2981 & -0.3064 & -0.3106 & -0.3133 & -0.3201 & -0.3214 & -0.3258 & -0.3217  \\
             0.7078 &  0.5409 &  0.1340 & -0.0076 & -0.0758 & -0.1137 & -0.1428 & -0.1687 & -0.1988 & -0.1992 & -0.2043  \\
            -0.3812 &  0.0551 &  0.8078 &  0.1571 &  0.1019 &  0.0468 &  0.0054 & -0.0585 & -0.2727 & -0.1935 & -0.2161  \\
             0.0772 & -0.1075 &  0.3832 & -0.2620 & -0.2674 & -0.2454 & -0.1940 & -0.1112 &  0.7612 &  0.0358 & -0.0681  \\
            -0.2859 &  0.3802 & -0.3067 &  0.2143 &  0.1633 &  0.1069 &  0.0393 &  0.1564 &  0.3987 & -0.2971 & -0.5679  \\
             0.4419 & -0.6531 &  0.0740 &  0.0669 &  0.1282 &  0.1292 &  0.1356 &  0.3206 & -0.0089 & -0.2228 & -0.4052  \\
            -0.0909 &  0.2057 &  0.0633 & -0.3272 &  -0.294 &  -0.212 & -0.0277 &  0.8243 & -0.1619 & -0.0144 &  0.0204  \\
             0.0078 & -0.0227 & -0.0088 &  0.0516 &  0.0335 &  0.0325 & -0.0011 &  0.0465 &  0.1350 & -0.8160 &  0.5552  \\
            -0.0184 &  0.1030 &  0.0274 & -0.6075 &  -0.075 &  0.2765 &  0.6906 & -0.2031 & -0.0002 & -0.1063 & -0.0880  \\
            -0.0026 &  0.0280 &  0.0228 & -0.3997 &  0.1912 &   0.674 & -0.5873 &  0.0538 & -0.0025 &  0.0143 &  0.0037  \\
        \end{array}\right] \left[\begin{array}{l}
            M_{\textit{FUV}} \\
            M_{\textit{NUV}} \\
            M_{\textit{u}} \\
            M_{\textit{g}} \\
            M_{\textit{r}} \\
            M_{\textit{i}} \\
            M_{\textit{z}} \\
            M_{\textit{Y}} \\
            M_{\textit{J}} \\
            M_{\textit{H}} \\
            M_{\textit{K}} \\
        \end{array}\right]
        $}
    \end{equation}
    \end{figure*}
    
    The 3rd component that represents $\sim$2\% of the variance is interesting. We show the distribution of galaxies in the D$_3$ concerning D$_2$ in Figure \ref{fig:D2-D3}. It is observed that there is an end in the transition around D$_2$ $\approx$2, which happens to be where the green valley galaxies lie. Therefore, we understand that galaxies bend in the multi-dimensional luminosity space when transitioning from star-forming to quiescent. Such non-linear structures should be handled with non-linear dimensionality reduction, which we consider as a comparison in Section \ref{sec:discussion_SOM}. Another actively researched area is manifold learning. There are now many available techniques such as t-SNE \citep[t-Distributed Stochastic Neighbor Embedding;][]{van-der-Maaten_2008} and UMAP \citep[Uniform Manifold Adaptation and Projection;][]{McInnes_2018}. An extension of this work with manifold learning is described in Takeuchi et al. in preparation.
    
\subsection{Interpretation of the axes} \label{sec:discussion_interpretation}

    A key goal in this work is to identify in a data-driven way the "fundamental" parameters that may be crucial for galaxy evolution. Thus, we discuss the intuition behind the two identified two parameters of the manifold.  

\begin{figure*}
        \centering
        \includegraphics[width=\linewidth]{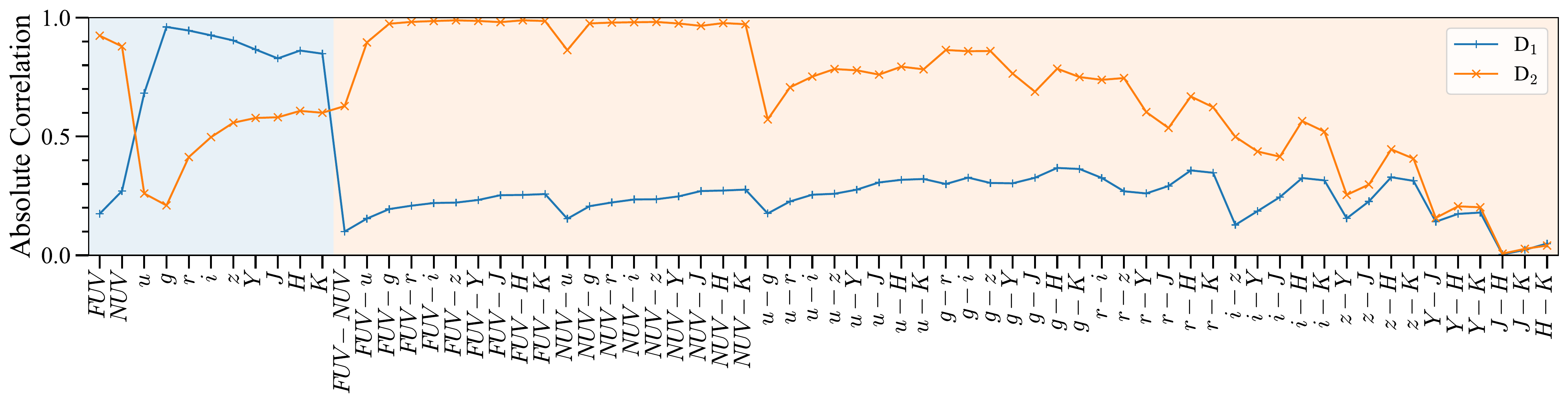}
        \caption{Absolute correlation between the manifold axes and the observable quantities. Considered observable quantities are the absolute magnitude and their color permutations. D$_1$ and D$_2$ are the most correlated with $g$-band magnitude and $FUV-z$ color.}
        \label{fig:correlation_color_luminosity}
    \end{figure*}

    Firstly, we identify the relationship between the manifold axes and the observables. We consider the original 11 luminosities and the colors derived from those 11 bands. In this case, we have 11 choose 2 = 55 colors in addition to the 11 bands. We show in Figure \ref{fig:correlation_color_luminosity} the absolute correlation between the 11 magnitudes + 55 colors = 66 features and the manifold axes. It is apparent that D$_1$ is correlated more with the bands (particularly $g$-band), and D$_2$ is correlated most with the UV-optical/UV-IR colors. Essentially, our manifold axis agrees with the traditional color-magnitude diagrams that use UV-optical colors with optical bands \citep[e.g.,][]{Strateva_2001,Blanton_2003} and UV-IR colors with optical bands \citep[e.g.,][]{Chilingarian_2012}. This result does not surprise us, as our technique and astronomers have identified the same parameters informative of observed galaxies over the years. It should be noted that \citet{Wild_2014} finds that their analysis of SEDs provided similar results to the traditional color-color diagrams ($UVJ$) instead of color-magnitude. 
    
    \cite{Conselice_2006} argued for a galaxy classification system where nearby galaxy properties could be expressed with three parameters, namely (1). mass or scale of a galaxy (2). recent star formation (3). Interaction degree. While we have not considered interaction in our analysis (Section \ref{sec:discussion_2modes}), we discuss our results with the first two parameters. Despite some dependence on D$_2$, D$_1$ found in this work has a close relationship with mass/scale as shown in Figure \ref{fig:manifold_SF}. D$_2$ is shown to be more related to the specific star formation rate than the star formation rate, essentially showing more of the evolutionary stage of the galaxy. The conclusion that D$_2$ is related to a galaxy's evolutionary stage is supported by our analysis of the manifold evolution, where modeled galaxies moved up along the D$_2$. 
    
    \citet{Eales_2018} with Herschel data has also argued for an sSFR vs. galaxy mass plane populated by a single "Galaxy Sequence", which is in complete agreement with our result. An essential discussion is that the "green valley" population, which resides between the star-forming and quenched populations, is an observational artifact. The reason is that galaxies with very low real SFR values have high uncertainty and form an artificial quenched population. A similar discussion was made in \citet{Feldmann_2017}, where the bimodality of the galaxy population in SFR was questioned. They argued that if one excludes the "dead" galaxies with zero SFR, the intrinsic distribution of galaxies in Log SFR is unimodal. Though we do not explicitly advocate for a unimodal solution, our analysis of luminosities has produced a single continuous manifold that galaxies evolve on. Thus, we favor the view that galaxies are not comprised of two distinct populations but one population which forms a manifold and that the evolutionary stage of a galaxy can be expressed by its location on this 2D manifold. 
    
\subsection{Two modes of evolution on the manifold} \label{sec:discussion_2modes}

    %dot product
    The result in Sections \ref{sec:Evolution-constSFR} and \ref{sec:Evolution-declSFR} suggests that there are two modes of evolution for galaxies (constant and exponential declining SFR). These modes suggest a close link between the evolutionary stages of galaxies and the gas accretion onto them, i.e., gas inflow maintains constant SFR or depletes the gas reserves when inflow is shut off, declining the SFR exponentially. This type of understanding between gas accretion and the evolution of SFR is incorporated in analytic models \citep[e.g.,][]{Bouche_2010,Dekel_2013,Lilly_2013,Dekel_2013,Forbes_2014b}, semi-analytical models \cite[e.g.,][]{Dutton_2010,Dave_2012,Mitchell_2014,Lacey_2016,Lagos_2018}, and hydrodynamic simulations \citep[e.g.,][]{Dave_2011,Vogelsberger_2014,Schaye_2015,Nelson_2018}.
    
    Due to our analysis's nature, we can only capture the longer timescale $\sim 1$ Gyr) evolution as we have focused on the mean evolutionary trajectories. However, many interesting and important physical processes such as galaxy mergers, galactic winds, and environmental effects are known to cause shorter timescales (<1 Gyr) variation of star formation \citep[e.g.,][]{Hernquist_1989,Mihos_1996,Roberts_1994,Oppenheimer_2008,McQuinn_2010,Sparre_2017,Torrey_2018, Wang_2020}. Despite this, galaxies in star-forming phase can be considered to be in quasi-steady state with extended periods of sustained star formation \citep[e.g.,][]{Daddi_2010,Genzel_2010,Dave_2012,Lilly_2013}. Besides, galaxy merging seems to only have a limited effect on SFR \citep[e.g.,][]{Noeske_2007a,Rodighiero_2011, Ellison_2013,Knappen_2015,Silva_2018,Pearson_2019}. Therefore, on longer timescales where quenching occurs, the identified two evolutionary modes or their combinations appear to be an accurate characterization.

\subsection{Predicting physical properties using the manifold}  \label{sec:discussion_property_predict}

    To demonstrate that the two manifold axes represent the galaxy properties well, we explore the performance of recovering the physical properties given the two axes. The recovery of SFR and $M_*$ with D$_1$ and D$_2$ is considered. We use the extra-trees regressor \citep{Geurts_2006} as implemented in \textsc{Python Sklearn}, which is similar to the popular random forest regression \citep{Bonjean_2019}. Despite the similarity, extra-trees have less overfitting by selecting the decision boundaries randomly. Ensemble methods like the above take the average over many estimators reducing the sample bias, and the forest of randomized tree methods like the above provide much more flexible approximations that are not possible by analytic functions such as polynomials. Though polynomials are easily expressed, in this case, we use the complex and flexible model to focus on the recoverability of the physical properties of the manifold without being biased by the model constraints. 
    
\begin{figure}
        \centering
        \includegraphics[width=\linewidth]{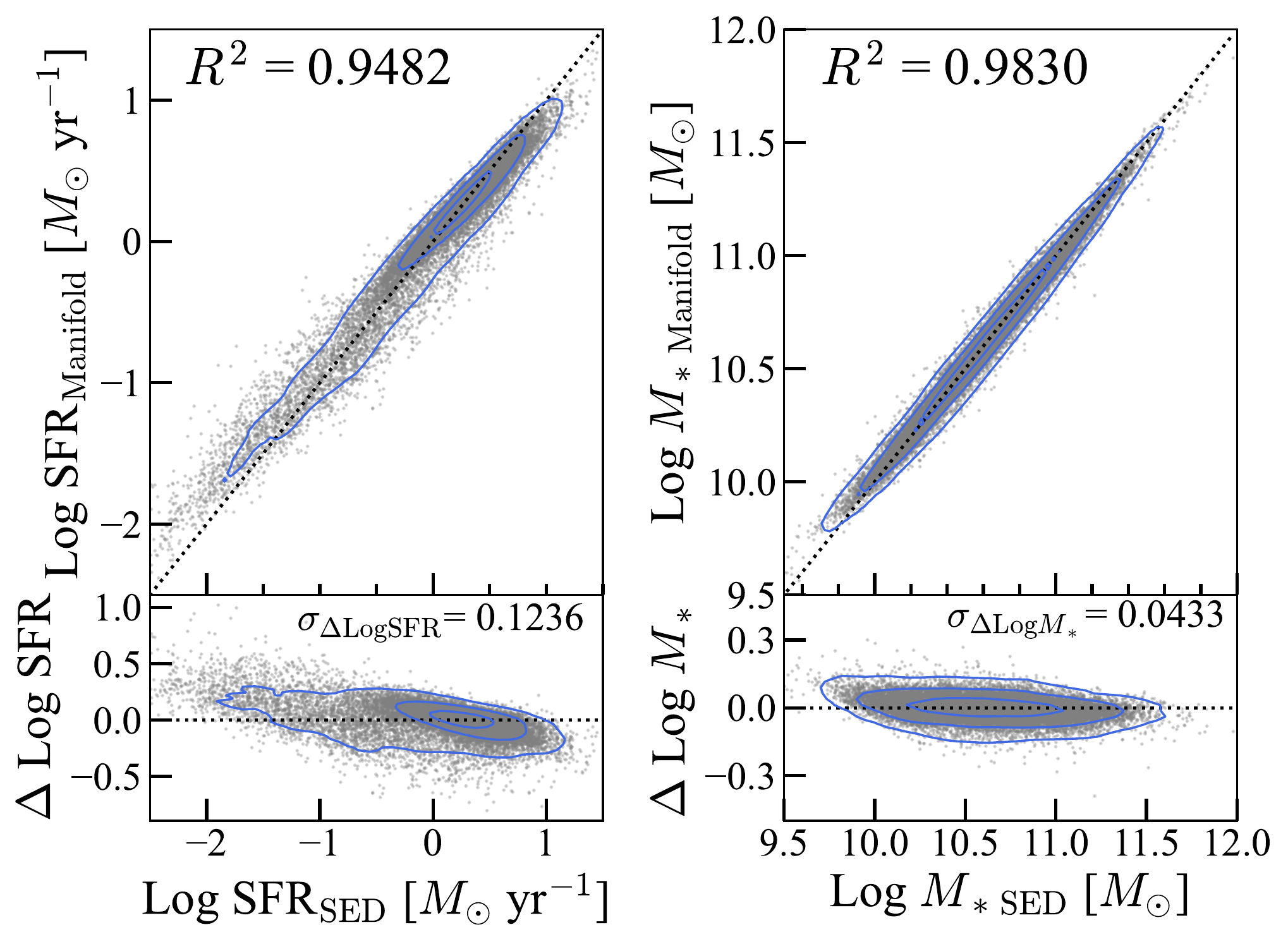}
        \caption{Prediction accuracy of star formation rates and stellar mass with the two manifold axes. Both properties are well recovered ($R^2>$ 0.94) by the galaxy manifold, which suggests that the manifold is a good representation of galaxies and their properties.}
        \label{fig:property_predict}
    \end{figure}

    Figure \ref{fig:property_predict} shows the accuracy of predicting SFR (right panel) and $M_*$ (left panel) on the manifold by comparison to the estimates using SED fitting considered to be the "truth". The SED-determined values give the abscissa, and the ordinate shows the predicted values using the extra-trees regression. Despite the slight bias in smaller values of both SFR and $M_*$, the predicted values mostly agree with the SED predicted values as shown with the coefficient of determination $R^2$. The prediction difference for SFR ($\Delta$ Log SFR = Log SFR$_{\mathrm{predicted}}$ - Log SFR$_{\mathrm{truth}}$) is within $\sim 0.25$ and for $M_*$ ($\Delta$ Log $M_*$ = Log $M_*$$_{\mathrm{predicted}}$ - Log $M_*$$_{\mathrm{truth}}$) is within $\sim 0.1$. The standard deviation of the prediction difference is $\sigma_{\Delta \mathrm{Log \ SFR}}=0.1236$ and $\sigma_{\Delta \mathrm{Log \ M_*}}=0.0433$. The lower SFR values are not well predicted compared to higher SFR galaxies. We argue that the poor predictability is attributed to the uncertainties of SED fitting and the poor relation between colors and sSFR below a certain threshold \citep{Eales_2017}.

\subsection{Recovering the Manifold axes using incomplete features} \label{sec:discussion_recover}

    In standard photometric observations, it is not guaranteed to have the complete 11 $K$-corrected magnitudes that were used for training. We explore the scenario when one has the observed magnitudes for all 11 bands or a subset of bands with redshift. This way, we can recover the manifold axes without the explicit need for $K$-correction.
    
\begin{figure}
        \centering
        \includegraphics[width=\linewidth]{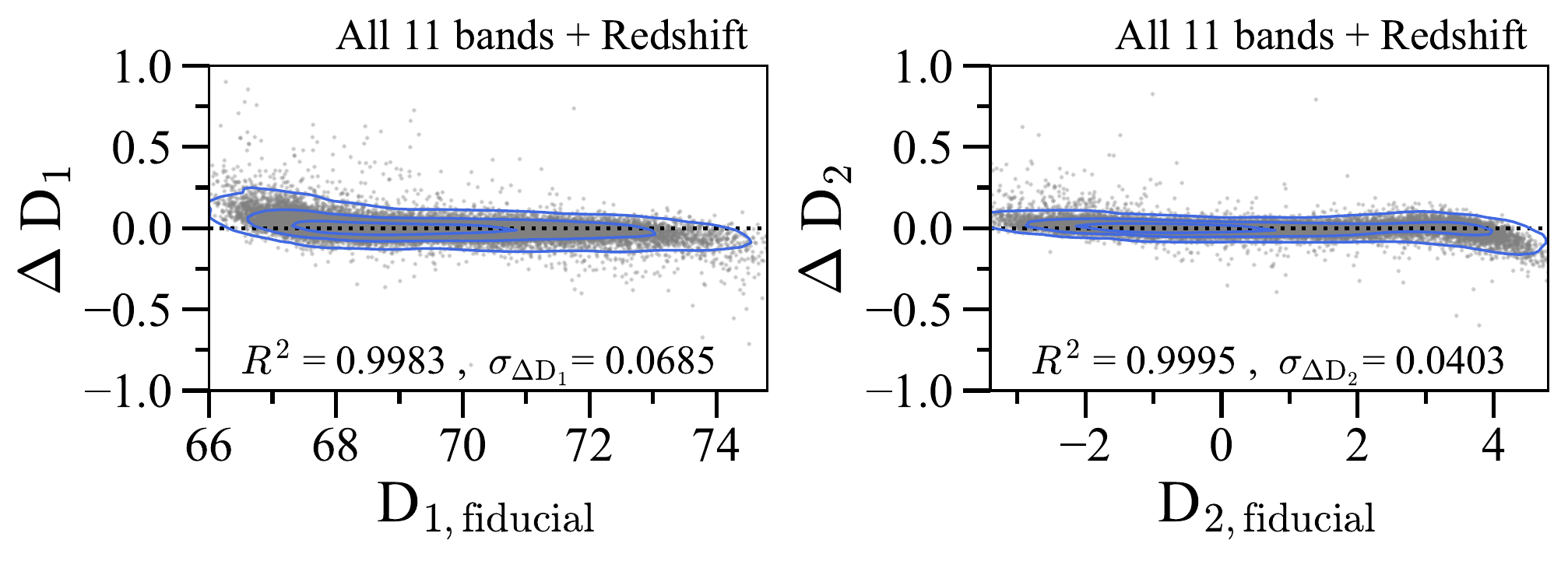}
        \includegraphics[width=\linewidth]{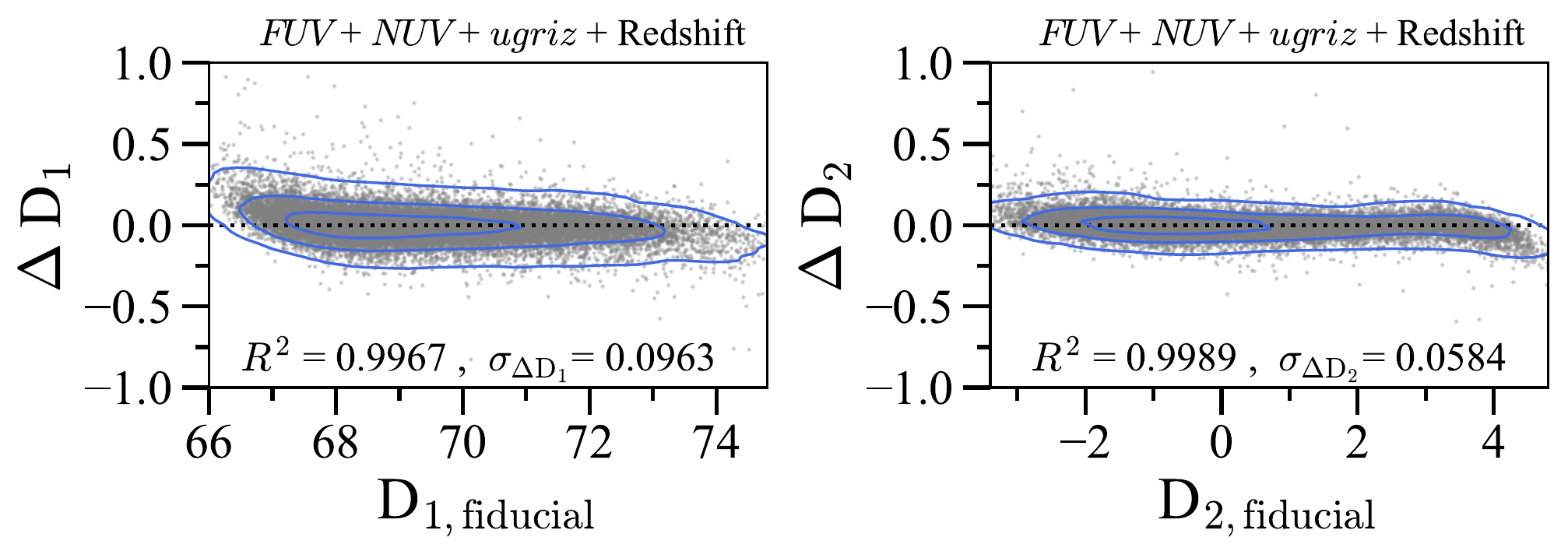}
        \includegraphics[width=\linewidth]{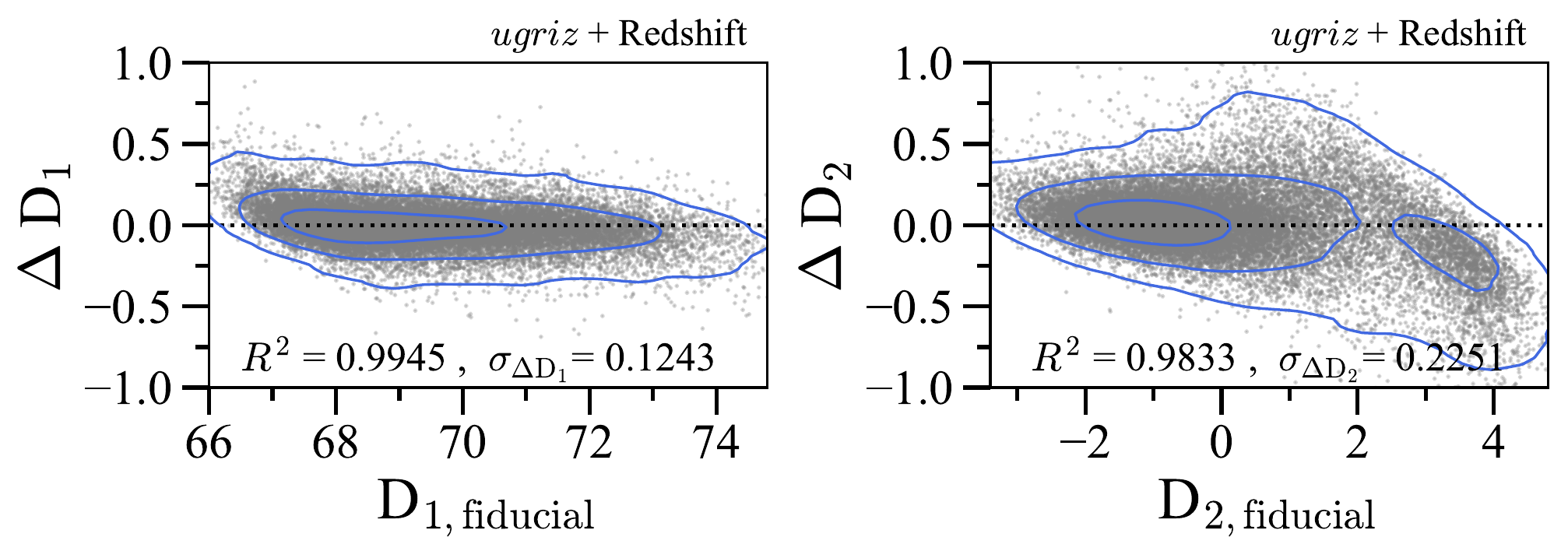}
        \caption{Residual between the predicted and true manifold parameters using incomplete features. The left/right columns show the residuals for D$_1$/D$_2$. The top row corresponds to the case when apparent magnitudes of all 11 bands are used with redshift. Middle row and bottom rows correspond to the estimation using $FUV$ + $NUV$ + $ugriz$ + redshift, and $ugriz$ + redshift, respectively. The residuals increased when the number of features used decreased.}
        \label{fig:predict_nonk_z}
    \end{figure}

    Three scenarios are considered in which a full or subset of bands with redshift is used to estimate the manifold axes. Firstly, we consider all 11 bands with redshift. For the other two cases, we consider the UV ($FUV$, $NUV$) and optical ($ugriz$) bands with redshift and optical ($ugriz$) bands with redshift. We use the same technique of extra trees as Section \ref{sec:discussion_property_predict} for the approximation. The residuals of D$_1$ ($\Delta$ D$_1$ = D$_{1 \mathrm{, predicted}}$ - D$_{1 \mathrm{, truth}}$) and D$_2$ ($\Delta$ D$_2$ = D$_{2 \mathrm{, predicted}}$ - D$_{2 \mathrm{, truth}}$) for the three cases are shown in Figure \ref{fig:predict_nonk_z}. $\Delta$ D$_1$ and $\Delta$ D$_2$ are shown in left and right panels. From the top row; all 11 bands + redshift, $FUV$ + $NUV$ + $ugriz$ + redshift, and $ugriz$ + redshift. 
    
    When all 11 bands are available with redshift, we can recover the manifold axes almost perfectly with $R^2$=0.9989. There is, however, some bias towards the lower D$_1$ and higher D$_2$, which corresponds to the quiescent galaxies. In the second row, we find that we can still recover to a large extent when the \textit{GALEX} $FUV$ and $NUV$ bands are included. The recoverability deteriorates when the above bands are removed, and only the optical $ugriz$ bands and redshift is used (third row). In this third case, D$_1$ is better reproduced than D$_2$, which is highly correlated with the UV-optical color.
    
\subsection{Comparison with a non-linear dimensionality reduction method} \label{sec:discussion_SOM}

\begin{figure}
        \centering
        \includegraphics[width=\linewidth]{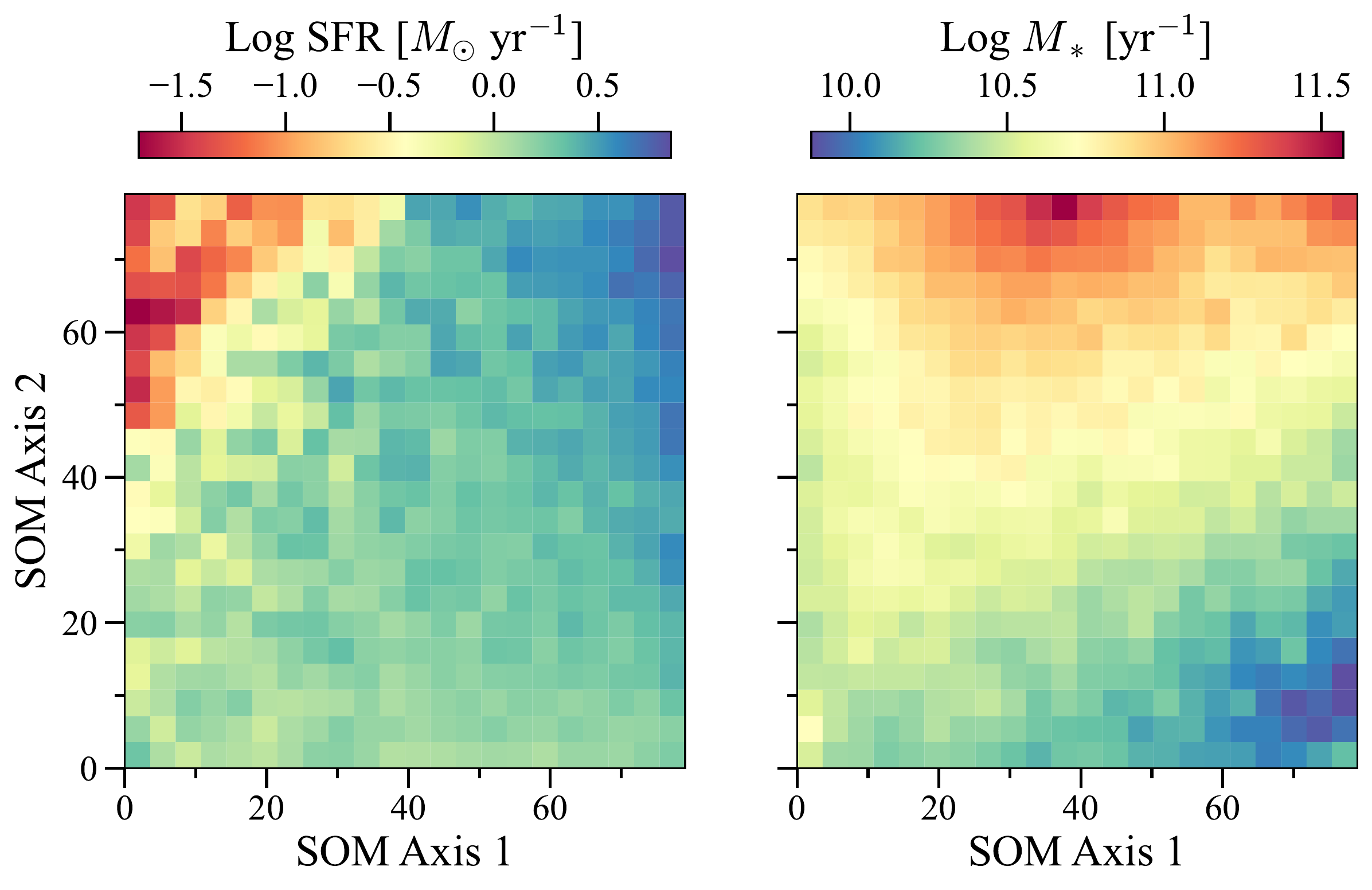}
        \caption{The distribution of star formation rates (SFR) and stellar masses ($M_*$) on the self-organizing map learned manifold.}
        \label{fig:SOM}
    \end{figure}

    For comparison with the linear method shown, we apply a popular non-linear dimensionality reduction called self-organizing maps \citep[SOM, ][]{Kohonen_2001}. SOM has been widely applied to galaxy observations  \citep[e.g.,][]{Miller_1996, Naim_1997, in_der_Au_2012, Rahmani_2018, Hemmati_2019, Davidzon_2022}. We use the parallelized implementation called \textsc{xPySom} \citep{xPySom} of the popular SOM implementation in python called \textsc{MiniSom} \citep{MiniSom}. We apply to the same 11 band data with 80 x 80 cells configuration, as in \citet{Davidzon_2022}. Figure \ref{fig:SOM} shows the result of the SOM grid with SFR and $M_*$ values. SOM has successfully found a manifold that is qualitatively similar to the galaxy manifold in that we can observe SFR gradients and $M_*$ on the SOM map. SOM SFR map shows the clear separation between the star-forming blue galaxies and the quiescent red galaxies. Similarly, SOM $M_*$ map has distributed the heavier and light galaxies to the top and the bottom, respectively. 
    
    However, the issue with non-linear dimensionality reduction methods is that the transformation from the data space to the SOM map is not trivial. SOM results of \citet{Davidzon_2022} show very complicated boundaries between quiescent galaxies, whereas, in our galaxy manifold, the boundary is a straight line on the manifold that achieves $\sim$ 85\% classification accuracy.
    Additionally, \citet{Holwerda_2022} shows that in their SOM result, kpc scale morphological features discussed above (Figure \ref{fig:manifold_morph}) cannot be well separated, unlike our galaxy manifold, which we consider to be a result of non-trivial mapping by the algorithm. While powerful, we believe reliance on non-linear methods can produce less predictive and unintended consequences in our applications.
    
    Additionally, since SOM is not a generative model, unless we calibrate the SOM grid \textit{aposteriori}, we can not generate new data measurements from the manifold. Though we do not demonstrate in this work, non-linear generative models such as variational autoencoders \citep[e.g.,][]{Portillo_2020} can be a powerful tool to simultaneously learn the complex underlying data structure and be generative models.
    
\subsection{Limitations of this Work} \label{sec:discussion_limitations}

    Our target in this work was to identify the manifold representing the physical properties of galaxies. In these types of works, the hope is that the data drives the physical interpretation of the data. However, that means the learned information is defined by the data we use for training. The dependence on the training sample also implies that modeling biases may also affect our results. 
    %Since inputs were $K$-corrected magnitudes, uncertainties of $K$-corrections provided in the catalogs (Section \ref{sec:data}) would inevitably be included. 
    
    The most critical obstacle, in this case, is the Malmquist bias \citep{Malmquist_1922}. However, the incompleteness is difficult to quantify, especially when data of multiple bands from multiple instruments are employed. Therefore, it is very challenging to determine what is truly representative. We tried to negate the issue through the volume-limited selection. However, the volume-limit censors the sample, reducing the dynamic range of the included properties in the manifold. In this work, the cut resulted in significantly removing high-SFR (Log SFR $\gtrsim$ 1 [$M_{\odot}$ yr$^{-1}$]), low-$M_*$ (Log $M_*$ $\lesssim 10$[$M_{\odot}$]), and very high-$M_*$ (Log $M_*$ $\gtrsim 11.5$ [$M_{\odot}$]). Weighting the samples without censoring may be advantageous in this case \citep{Takeuchi_2000}, which will be considered in the future. Alternately, training on simulations will overcome the dependency on sample selection. However, the issue then shifts to the accuracy of the modeling as even the state-of-the-art simulations still fail to precisely reproduce the observed color distributions \citep[e.g.,][]{Nelson_2018}.
    
    An unavoidable consequence of dimensionality reduction is the loss of possibly critical information. While the two axes can contain the most information about the overall evolutionary stages of galaxies, there may be additional information that may help understand galaxies better encoded in higher dimensions. Additionally, using photometry instead of spectroscopy also inherently limits the attributes the manifold can express. Details of the interstellar medium or the active galactic nuclei often require the line emission data to decipher correctly \citep[e.g.,][]{Baldwin_1981,Kewley_2002,Kauffmann_2003,Brinchmann_2008}. The representations of galaxies with the two axes will degenerate when required to be projected to alternate spaces. For example, the morphological classification provided in Section \ref{sec:physical_properties_morph} is not a clear boundary that separates the classes. Additional information, such as line diagnostics, may provide better class separation.

    Additionally, we have not considered measurement errors in our analysis. Points to consider would be the sample selection and the dimensionality reduction. As for the sample selection, some galaxies may have been included or left out due to photometric errors. Each band also has heterogeneous errors making the consideration very complicated. SVD, like many dimensionality reduction methods available today, cannot handle uncertainties out of the box. While proper treatment of uncertainties with Monte Carlo sampling may be possible, we deem the difference insignificant and that it will not make a qualitative difference in the result presented in this work.

\section{{Conclusion}} \label{sec:conclusion}

    Here we reported the discovery of two parameters that define the local galaxy distribution within the multi-dimensional luminosity space from far ultraviolet to near-infrared wavelengths. Analytic linear transformations relate the two parameters found by dimensionality reduction of the observable luminosities. These two parameters then define a "galaxy manifold", where galaxies exist and evolve on the manifold. The found manifold can be considered the ideal representation of the galaxy distribution in the color-magnitude space and provides a convenient tool to characterize galaxies.

    The existence of two parameters representing 93.2\% of the information of our galaxy sample at redshift<0.1 suggests that the galaxy manifold derived here is likely to be one of the best representations of galaxy physical parameter space. Such a manifold provides tremendous potential for future studies. Firstly, such representations will give efficient boundaries for galaxy classification tasks as explored in Section \ref{sec:physical_properties_morph}. Similarly, galaxies represented on a manifold can be helpful for clustering analysis \citep[e.g.,][]{Siudek_2018, Turner_2019,Yesuf_2020} to find subpopulations that can reveal details on the various evolutionary trajectories galaxies take. 
    
    A vital aspect of this work is the consideration of evolution on the manifold. On the 2D manifold, we show two modes of star formation histories almost orthogonal to each other, which can be a basis for any complex star formation history. By calibration of the manifold, we also show evolution tracks derived using simple analytic equations of evolution involving gas and star formation. Assumed parameters are minimal and reproduce consistent SFHs. Parameterization in terms of D$_1$ and D$_2$ allows deriving the evolution of any physical parameter on the manifold. To our knowledge, this work is the first to consider the evolution of galaxies parameterized latent space that is learned in an unsupervised manner.
    
    For more practical applications, a learned manifold can aid in finding solutions for unconstrained problems such as photometric redshift estimation \citep[e.g.,][]{Carrasco-Kind_2014, Speagle_2017}. Similarly, a low dimensional space where the galaxies lie suggests that we do not have to search the entire parameter space, leading to more efficient estimation of physical properties from observations \citep{Hemmati_2019, Davidzon_2022}. The upcoming projects like Legacy Survey of Space and Time \citep[LSST;][]{Ivezic_2019} and Euclid \citep{Laureijs_2011} will produce unprecedented amounts of data that needs to be handled efficiently to achieve their scientific goals. Pipelines will benefit the inference accuracy and speed when the galaxy manifold is incorporated as prior knowledge, requiring much fewer (two) free parameters.
    
    While we have left out the redshift evolution of the manifold for future papers, constraining and calibrating such manifolds at high redshifts may prove imperative to understanding galaxies' formation and evolution at those epochs. Such will be more informative than summary statistics such as mass or luminosity functions. Similar to \citet{Masters_2015}, we will consider the ways of efficiently sampling the parameter space in both the magnitude/colors and the physical properties to constrain the manifold in future work.
    
    The concept of manifolds has received a recent resurgence of interest, especially in the data-driven science community. While we have succeeded in understanding and simulating complex processes that galaxies undergo to a large extent, there remain crucial unanswered questions. Data-driven science with the latest data and methods may hold the key to answering those questions. We strongly believe that galaxy manifolds will be a robust framework in the future to both characterize and also understand galaxy evolution across cosmic time.

\section*{Acknowledgment}

    % First, we thank the anonymous referee for her/his careful reading of the manuscript to provide suggestions that signiﬁcantly improved this paper. 
    SC is supported by the Japan Society for the Promotion of Science (JSPS) under Grant No. JP21J23611.
    This work has been supported by JSPS Grants-in-Aid for Scientific Research  (TT: JP19H05076 and DK: JP21K13956). This work has also been supported in part by the Sumitomo Foundation Fiscal 2018 Grant for Basic Science Research Projects (180923), and the Collaboration Funding of the Institute of Statistical Mathematics “New Development of the Studies on Galaxy Evolution with a Method of Data Science”. 

\section*{Data availability}

The data underlying this article is publicly available at \hyperlink{http://rcsed.sai.msu.ru}{http://rcsed.sai.msu.ru}. Any derived data products will be provided on request. 

%%%%%%%%%%%%%%%%%%%%%%%%%%%%%%%%%%%%%%%%%%%%%%%%%%

%%%%%%%%%%%%%%%%%%%% REFERENCES %%%%%%%%%%%%%%%%%%

\bibliographystyle{mnras}
\bibliography{0-references}

%%%%%%%%%%%%%%%%%%%%%%%%%%%%%%%%%%%%%%%%%%%%%%%%%%

%%%%%%%%%%%%%%%%% APPENDICES %%%%%%%%%%%%%%%%%%%%%

\appendix
\section{Redshift Distribution on the manifold} \label{sec:redshift}

\begin{figure}
        \centering
        \includegraphics[width=\linewidth]{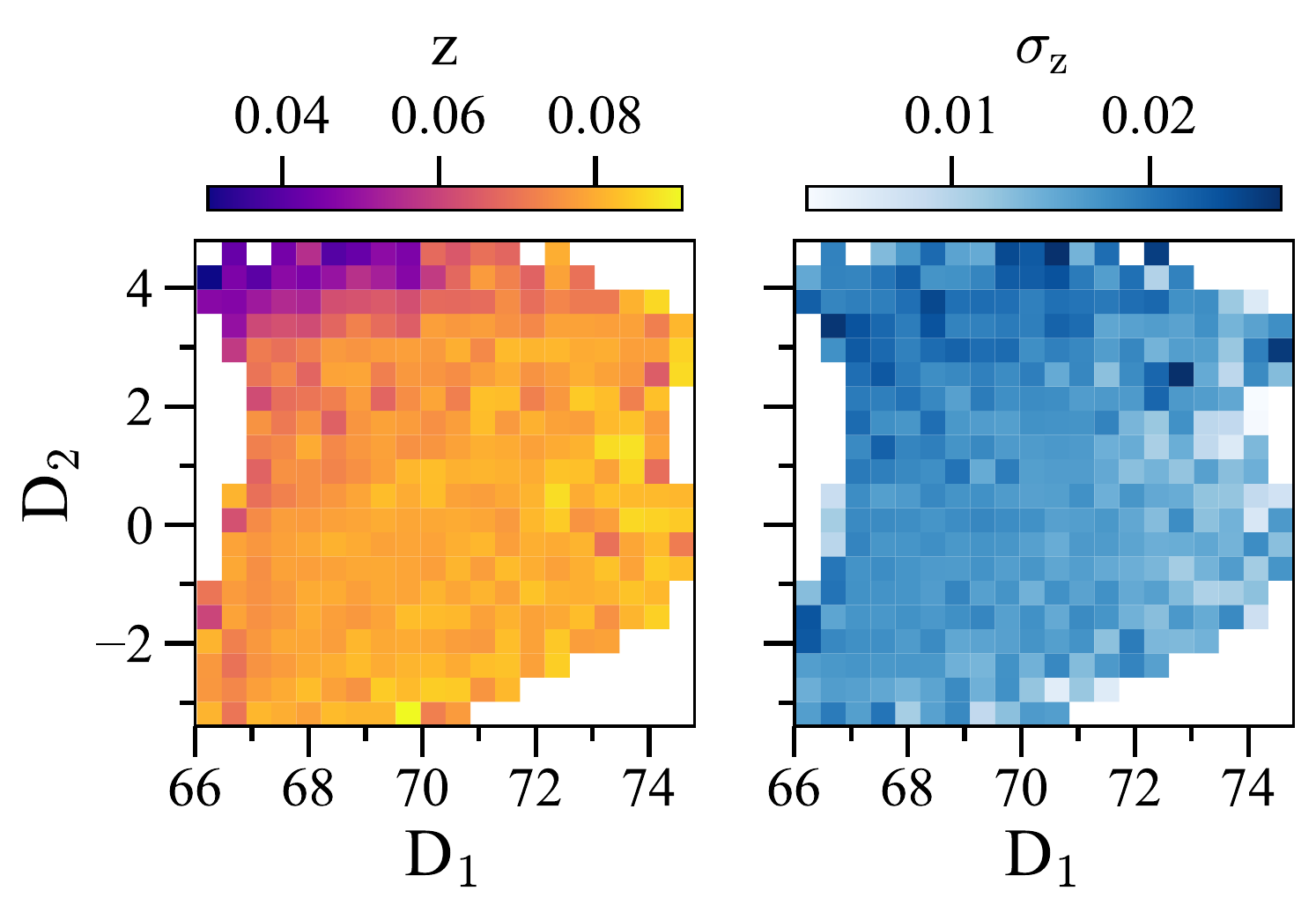}
        \caption{The distribution of redshift on the galaxy manifold. The left panel shows the median vales and right panels shows the scatter. For majority of the region, the median redshift is constant.}
        \label{fig:manifold_redshift}
    \end{figure}

    We show the median and standard deviation of the redshift distribution on the manifold in Figure \ref{fig:manifold_redshift}. As expected due to our volume limit of the sample, the manifold is mostly redshift independent except for a small region in upper left. 
    
\section{Parameterization of galaxy properties on the manifold} \label{sec:fitting}

    We showed in Section \ref{sec:discussion_property_predict} that galaxy physical properties can be well expressed as a function of the manifold axes. We provide parameterization for the physical properties discussed in this work in the form of fitted coefficients for the 2D polynomial of second order. The fitted models for various properties allows one to estimate those properties with the knowledge of D$_1$ and D$_2$. The fitting was done using robust linear regression implemented in \textsc{python statsmodels}. Use of robust linear regression prevents the fits to be affected by any outliers. Particularly, we use M-estimators that minimizes the function,
    \begin{equation}
        Q\left(e_{i}, \rho\right)=\sum_{i} \rho\left(\frac{e_{i}}{s}\right),
    \end{equation}
    where $e_i$ are the residuals, and $s$ is a scale, and $\rho$ is the function that minimizes the effect of outliers. The weighting used was Huber’s T norm with the median absolute deviation scaling. The fitted coefficients are provided in Table \ref{table:polynomial_coeff}. As seen by the coefficients, first order polynomial with the interaction term is mostly sufficient to approximate the distribution.
    
    \begin{table*}
    \begin{tabular}{|l|c|c|c|c|c|c|c|}
    \hline 
     & \textbf{Log SFR} & \textbf{Log $M_*$} & \textbf{Log sSFR} & \textbf{Log $M_{H_I}$} & \textbf{Log $M_{H_2}$} & \textbf{Log $M_{\mathrm{gas}}$} & \textbf{T-type}\\ 

     & \textbf{[$M_{\odot}$ yr$^{-1}$]} & \textbf{[$M_{\odot}$]} & \textbf{[yr$^{-1}$]} & \textbf{[$M_{\odot}$]} & \textbf{[$M_{\odot}$]} & \textbf{[$M_{\odot}$]} & \\ 
    \hline \hline
    
    $\beta_0$ & -8.660 $\pm$ 2.244 &  4.118 $\pm$ 0.911 & -17.304 $\pm$ 2.848 &  27.777 $\pm$ 1.625 &  9.997 $\pm$ 1.505 &  22.300 $\pm$ 1.693 &  94.951 $\pm$ 13.667\\
    $\beta_1$ &  0.164 $\pm$ 0.064 &  0.040 $\pm$ 0.026 &  0.255 $\pm$ 0.081 & -0.599 $\pm$ 0.046 & -0.118 $\pm$ 0.043 & -0.451 $\pm$ 0.048 & -2.668 $\pm$ 0.391\\
    $\beta_2$ & -0.220 $\pm$ 0.037 &  0.403 $\pm$ 0.015 & -0.644 $\pm$ 0.047 & -0.241 $\pm$ 0.027 & -0.069 $\pm$ 0.025 & -0.252 $\pm$ 0.028 &  1.542 $\pm$ 0.227\\
    $\beta_3$ & -0.001 $\pm$ 0.000 &  0.001 $\pm$ 0.000 & -0.002 $\pm$ 0.001 &  0.005 $\pm$ 0.000 &  0.002 $\pm$ 0.000 &  0.004 $\pm$ 0.000 &  0.019 $\pm$ 0.003\\
    $\beta_4$ &  0.001 $\pm$ 0.001 & -0.004 $\pm$ 0.000 &  0.005 $\pm$ 0.001 &  0.002 $\pm$ 0.000 & -0.000 $\pm$ 0.000 &  0.002 $\pm$ 0.000 & -0.038 $\pm$ 0.003\\
    $\beta_5$ & -0.039 $\pm$ 0.000 & -0.012 $\pm$ 0.000 & -0.029 $\pm$ 0.001 & -0.016 $\pm$ 0.000 & -0.027 $\pm$ 0.000 & -0.025 $\pm$ 0.000 &  0.041 $\pm$ 0.003\\
    Scatter & 0.235 &  0.095 &  0.308 &  0.168 &  0.151 &  0.171 &  1.413 \\
    \hline
    \end{tabular}
    \caption{Fitted coefficients for the galaxy properties with robust linear regression of second order polynomials as a function of D$_1$ and D$_2$. The equation is of the form: $Y = \beta_0 + \beta_1*\mathrm{D}_1 + \beta_2*\mathrm{D}_2 + \beta_3*\mathrm{D}_1^2 + \beta_4*\mathrm{D}_1\mathrm{D}_2 + \beta_5*\mathrm{D}_2^2$, where $Y$ is the galaxy property. The bottom row shows the scatter of the residuals.}
    \label{table:polynomial_coeff} 
    \end{table*}

\section{Third axis of the manifold} \label{sec:3D}

\begin{figure}
        \centering
        \includegraphics[width=\linewidth]{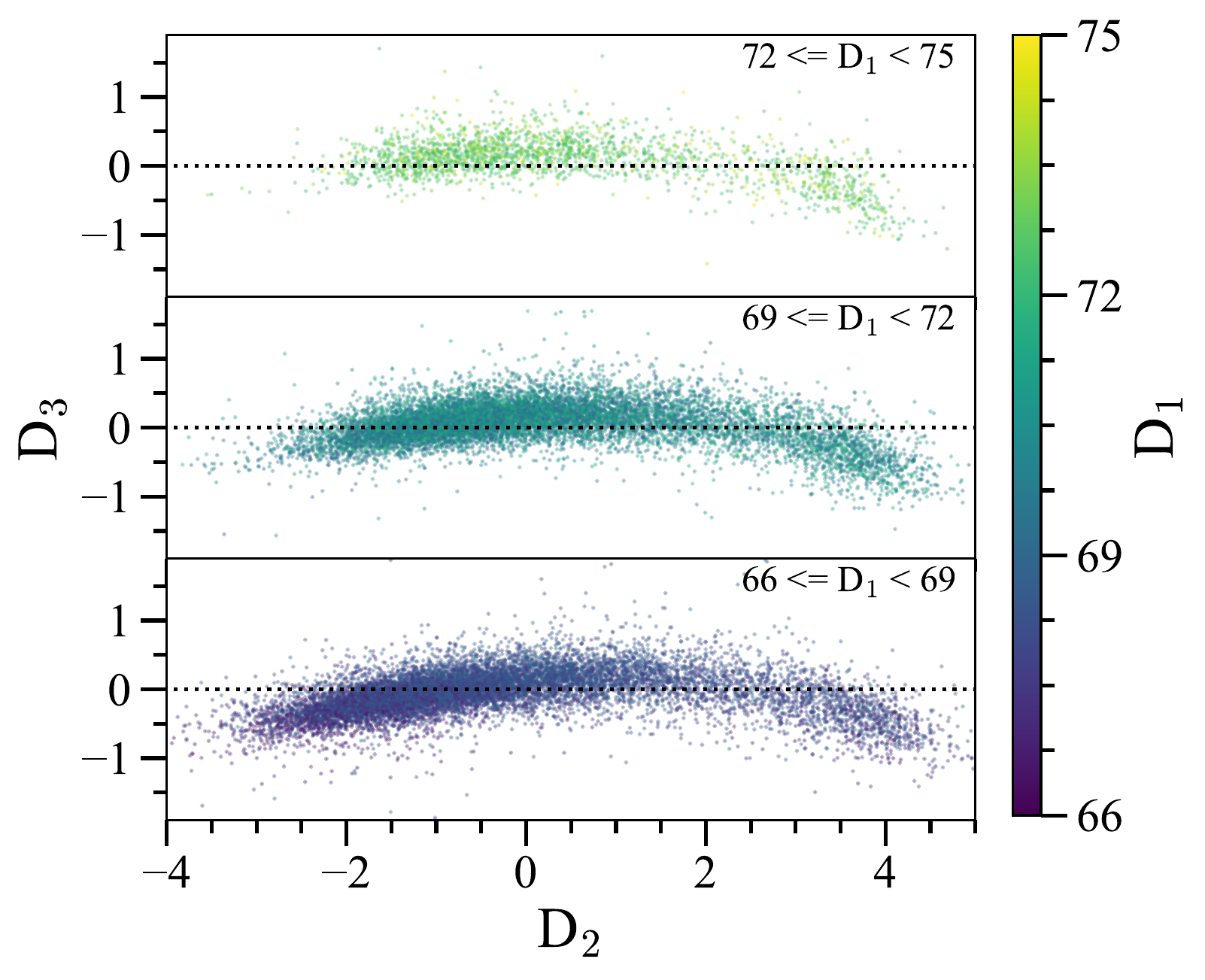}
        \caption{The distribution of galaxies in the D$_2$ vs. D$_3$ plane binned into three ranges in D$_1$. There is a slight bending of the manifold at the region between star-forming and quiescent galaxies, which is not expressed when considering only the first two parameters.}
        \label{fig:D2-D3}
    \end{figure}

    The first two axes explain 93.2\% of the variance in our galaxy sample (Section \ref{sec:data}). D$_3$ explain $\sim2$\% of the variance. We show the D$_2$ vs. D$_3$ plane in Figure \ref{fig:D2-D3}.
    
%%%%%%%%%%%%%%%%%%%%%%%%%%%%%%%%%%%%%%%%%%%%%%%%%%

% Don't change these lines
\bsp    % typesetting comment
\label{lastpage}
\end{document}